\documentclass[a4paper,aps,pra,reprint,showkeys,amssymb,floatfix,longbibliography,superscriptaddress,accepted=2025-10-19]{quantumarticle}

\pdfoutput=1
\usepackage{graphicx}
\usepackage{amsmath,amsfonts,amssymb}
\usepackage{hyperref}
\usepackage{braket}
\usepackage{xcolor}
\usepackage{tikz}
\usepackage{color}
\usepackage{enumerate}
\usepackage[caption=false]{subfig}
\usepackage{multirow}
\usepackage{qcircuit}
\usepackage[normalem]{ulem}
\usepackage{placeins}
\usepackage{xurl}
\usepackage{dsfont}
\usepackage{braket}
\usepackage[multidot]{grffile}
\usepackage{mathtools}
\usepackage{float}
\usepackage{textcomp}
\usepackage{siunitx}

\usepackage{amsthm}
\newtheorem{theorem}{Theorem}[section]

\usepackage{float}
\makeatletter
\let\newfloat\newfloat@ltx
\makeatother

\usepackage{algorithm}
\usepackage[noend]{algpseudocode}

\floatname{algorithm}{Protocol }

\algdef{SE}[DOWHILE]{Do}{doWhile}{\algorithmicdo}[1]{\algorithmicwhile\ #1}%

\newcommand\subsetsim{\mathrel{%
  \ooalign{\raise0.2ex\hbox{$\subset$}\cr\hidewidth\raise-0.8ex\hbox{\scalebox{0.9}{$\sim$}}\hidewidth\cr}}}

\newcommand{\equref}[1]{Eq.~(\ref{#1})}
\newcommand{\figref}[1]{Fig.~\ref{#1}}
\newcommand{\secref}[1]{Sec.~\ref{#1}}

\DeclarePairedDelimiterX{\abs}[1]{\lvert}{\rvert}{\ifblank{#1}{{}\cdot{}}{#1}}

\begin{document}

\title{Learning-Driven Annealing with Adaptive Hamiltonian Modification for Solving Large-Scale Problems on Quantum Devices}

\author{Sebastian Schulz}
\thanks{Corresponding author: Sebastian Schulz}
\email{se.schulz@fz-juelich.de}
\affiliation{J\"ulich Supercomputing Centre,
Institute for Advanced Simulation,\\
Forschungszentrum J\"ulich, 52425 J\"ulich, Germany}
\author{Dennis Willsch}
\affiliation{J\"ulich Supercomputing Centre,
Institute for Advanced Simulation,\\
Forschungszentrum J\"ulich, 52425 J\"ulich, Germany}
\affiliation{Faculty of Medical Engineering and Technomathematics, University of Applied Sciences Aachen,\\
52428 J\"ulich, Germany}
\author{Kristel Michielsen}
\affiliation{J\"ulich Supercomputing Centre,
Institute for Advanced Simulation,\\
Forschungszentrum J\"ulich, 52425 J\"ulich, Germany}
\affiliation{AIDAS, 52425 J\"ulich, Germany}
\affiliation{RWTH Aachen University, 52056 Aachen, Germany}

\date{2025-10-19}

\begin{abstract}
We present Learning-Driven Annealing (LDA), a framework that links individual quantum annealing evolutions into a global solution strategy to mitigate hardware constraints such as short annealing times and integrated control errors. Unlike other iterative methods, LDA does not tune the annealing procedure (e.g.~annealing time or annealing schedule), but instead learns about the problem structure to adaptively modify the problem Hamiltonian. By deforming the instantaneous energy spectrum, LDA suppresses transitions into high-energy states and focuses the evolution into low-energy regions of the Hilbert space. We demonstrate the efficacy of LDA by developing a hybrid quantum-classical solver for large-scale spin glasses. The hybrid solver is based on a comprehensive study of the internal structure of spin glasses, outperforming other quantum and classical algorithms (e.g., reverse annealing, cyclic annealing, simulated annealing, Gurobi, Toshiba's SBM, VeloxQ and D-Wave hybrid) on 5580-qubit problem instances in both runtime and lowest energy. LDA is a step towards practical quantum computation that enables today's quantum devices to compete with classical solvers.
\end{abstract}


\maketitle


\section{Introduction}
Combinatorial optimization problems (COPs) are ubiquitous in computer science, with important applications in finance~\cite{COP_Finance}, scheduling~\cite{COP_Scheduling}, machine learning~\cite{COP_ML_1, COP_ML_2}, computational biology~\cite{COP_Biology}, and operations research~\cite{COP_OperationsResearch}. For many such problems, finding the optimal solution is equivalent to finding the ground state of an associated Ising spin-glass system~\cite{ISING_Formulations}. The hardness of COPs is related to the presence of opposing spin interactions, leading to frustration and a glass phase. The latter is characterized by the presence of many low-energy local minima, that are separated by large energy barriers. This makes solving large-scale COPs often intractable on classical computers, quickly exceeding runtimes of $24$ hours.

With the advent of new generations of quantum annealers from D-Wave Systems Inc.~comprising more than $5000$ qubits, a promising approach is the use of quantum annealing (QA)~\cite{QA_Industry, QA_Supremacy, ThreeDWaveGenerations}. Inspired by the cooling of physical systems, QA uses quantum fluctuations caused by a transverse field to navigate the energy landscape of the spin system~\cite{QA_Overview}. If the quantum evolution is performed adiabatically, the system arrives in the ground state of the spin glass. However, if the evolution is too fast, the system scatters into higher energy states through a series of quantum phase transitions (QPTs, used in the sense of Refs.~\cite{QA_QPT_1,QA_QPT_2,QA_QPT_3,SG_QPT_Theory}). Hence, on real devices, finite annealing times and integrated control errors limit the performance, often making deep low-energy states (including the ground state) inaccessible~\cite{QA_fails_1, QA_fails_2}. To address this issue, several authors recently investigated iterative procedures, such as reverse annealing~\cite{IRA_Theory_1, IRA_Theory_2, IRA_Theory_3, IRA_Thermal_1} and cyclic annealing~\cite{Cyclic_annealing_1, Cyclic_annealing_2, Ra_Hg}, in which the annealing is repeatedly initialized from the best known classical state. In doing so, the hope is that QA gradually converges to the ground state, but in practice, these procedures often fail, as they become stuck in high-energy valleys~\cite{IRA_Failing_1, IRA_Failing_2}. 

We show that systematically learned modifications to the problem Hamiltonian itself can greatly improve the performance of iterative QA on noisy hardware. With Learning-Driven Annealing (LDA), we propose a framework to link individual QA runs into a global solution strategy to mitigate hardware constraints such as finite annealing times and integrated control errors~\cite{QPU_ICE}. LDA works by analyzing the states sampled from QA to adaptively modify the problem Hamiltonian to the information learned about the energy landscape. This energetically isolates low-energy valleys in the instantaneous energy spectrum, focussing subsequent annealing runs into a promising region of the Hilbert space. As a result, the QPU can reach deeper regions of the energy spectrum that are otherwise inaccessible. While we do not claim a quantum advantage per se, LDA enables current NISQ devices to tackle COPs more efficiently as long as they can be embedded onto their hardware.

We demonstrate the capabilities of LDA by designing a hybrid quantum-classical solver for large-scale spin glasses, that is based on alternating local and global search protocols. Using the D-Wave Advantage1 5.4 QPU~\cite{QPU_Advantage_5_4} in Jülich, Germany, we benchmark our hybrid solver against leading quantum and classical algorithms, including reverse annealing~\cite{IRA_Theory_1}, cyclic annealing~\cite{Cyclic_annealing_1, Cyclic_annealing_2}, simulated annealing using JUPTSA~\cite{JUPTSA}, Gurobi~\cite{Gurobi}, Toshiba's SBM~\cite{Toshiba_SBM}, VeloxQ~\cite{VeloxQ, VeloxQ2} and D-Wave hybrid~\cite{D_Wave_Hybrid}, on $5580$-qubit NAT-7~\cite{NAT_7_1, NAT_7_2} spin glasses. The results show that our hybrid solver outperforms all competing strategies in both runtime and lowest energy. LDA is a step towards practical quantum computation that enables today's quantum annealers to compete with classical solvers.

The remainder of this paper is organized as follows: In Sec.~\ref{sec:spin_glass}, we introduce the spin-glass formulation and provide insights into its underlying structures. Section~\ref{sec:qa} concerns adiabatic quantum evolutions and discusses its limitations on modern quantum annealers. In Sec.~\ref{sec:lda}, we introduce the LDA framework, showcasing how learned information from sampled states can be used to guide the annealing evolution by modifying the problem Hamiltonian. In Sec.~\ref{sec:hybrid_optimization}, we demonstrate how LDA can be used in advanced algorithms by designing a hybrid solver for spin glasses based on alternating local and global search protocols. Finally, in Sec.~\ref{sec:benchmarks}, we benchmark our hybrid solver against various classical and quantum solvers, followed by a summary of our findings in Sec.~\ref{sec:conclusion}.


\begin{figure*}
  \centering
  \includegraphics[width=\linewidth]{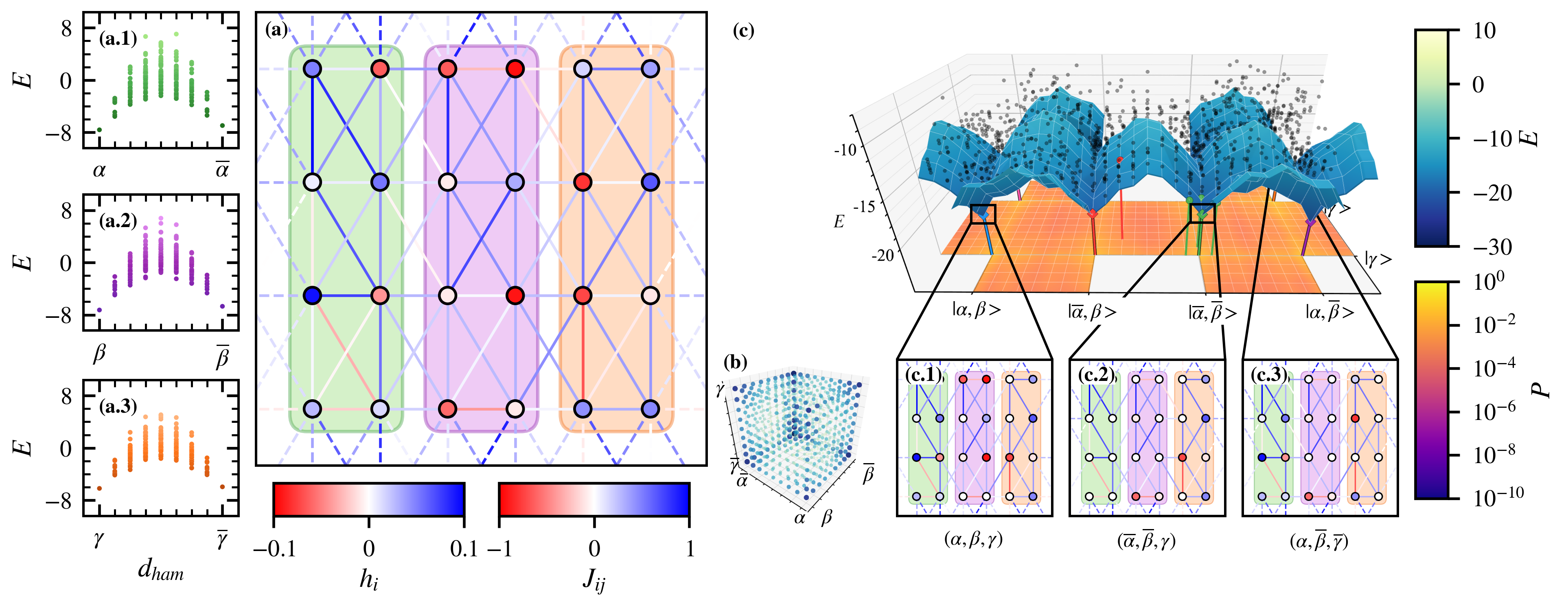}
  \caption{A $24$-qubit spin-glass with $N_{dom} = 3$. (a) Spin-glass graph with the biases $h_i$ as nodes and couplers $J_{ij}$ as edges. Dashed edges indicate periodic boundary conditions. The colored rectangles represent the spin domains: $\mathcal{D}_1$ (green), $\mathcal{D}_2$ (purple), and $\mathcal{D}_3$ (orange). Each domain has two distinct minima: $\alpha$ and $\overline{\alpha}$ ($\mathcal{D}_1$), $\beta$ and $\overline{\beta}$ ($\mathcal{D}_2$), $\gamma$ and $\overline{\gamma}$ ($\mathcal{D}_3$). Panels (a.1-3) show the eigenenergies $E$ of the corresponding domains as a function of the Hamming distance $d_{ham}$ to the respective ground state $\alpha, \beta, \gamma$. (b) Hypercube representation of the Hilbert space, with each axis representing the Hamming distance $d_{ham}$ from the ground state ($\alpha, \beta, \gamma$) of the respective spin domains $\mathcal{D}_k$. Color indicates the lowest eigenenergy at each point. The corners of the hypercube correspond to the $2^{N_{dom}} = 8$ local minima of the spin glass. (c) Energy landscape of the spin glass obtained by unfolding the hypercube facets. Each point is projected onto the closest facet (smallest Hamming distance), with the blue surface representing the lowest eigenenergies on these facets. The eight local minima are marked by colored diamonds at the facets' corners, with the states surrounding it being the energy valleys. Additionally, black dots indicate $2000$ samples from a quantum annealing simulation with annealing time $T = 10$, with the probability density shown on the lower orange plane. Panels (c.1-3) highlight the specific biases $h_i$ and couplers $J_{ij}$ satisfied by three of the eight local minima, with the unsatisfied terms set to $0$ (white). The color scheme matches that of panel (a).}
  \label{fig:spin_glass_graph}
\end{figure*}

\section{Spin-glass problems}
\label{sec:spin_glass}
Spin glasses (SG) are magnetic systems of quenched disorder, imposing conflicts between interacting magnetic moments~\cite{SG_Theory}. In the Edwards-Anderson model of SG~\cite{SG_Theory_EA}, $N$ Ising spins $\sigma^z_i$, $i\in\{0, \, \dots \, , \, N-1\}$, are placed on the sites of a regular graph, with the Hamiltonian defined as
    \begin{align}
        H = \sum_{i<j} J_{ij} \sigma^z_i \sigma^z_j + \sum_i h_i \sigma^z_i.
        \label{eq:H_EA}
    \end{align}
Here $J_{ij}$ with $i<j$ denotes the coupling strength between spins $i$ and $j$, and $h_i$ represents a locally applied external magnetic field bias on spin $i$. 

We consider both triangular lattices with periodic boundary conditions (see \figref{fig:spin_glass_graph}a) and the Pegasus graph used by the D-Wave Advantage1 5.4 quantum annealer~\cite{QPU_Advantage_5_4}. In the following, we choose $\abs{h_i}\ll \abs{J_{ij}}$ to ensure the generation of complex SG~\cite{SG_Hard_Instances}. Notably, for $h_i=0$, $H$ exhibits a global $\mathbb{Z}_2$ symmetry.

Throughout this study, we use Greek letters, e.g. $\alpha = \alpha_{N-1}\cdots\alpha_1\alpha_0$, to represent a generic eigenstate of \equref{eq:H_EA}, with $\alpha_i = \pm 1$ denoting the polarization of the $i$-th spin. We also use the notation $\alpha^* = \alpha_{N-1}^*\cdots\alpha_1^*\alpha_0^*$ to refer to a local (in Hamming distance) minimum of \equref{eq:H_EA}. The eigenenergy of a state $\alpha$ is given by $E_{\alpha} = \sum_{i<j} J_{ij} \alpha_i \alpha_j + \sum_i h_i \alpha_i$ and its bit string representation is written as $\texttt{a} = \texttt{a}_{N-1}\cdots\texttt{a}_1\texttt{a}_0$, with $\texttt{a}_i=(1+\alpha_i)/2$.

For a given state $\alpha$, we define the set of satisfied (i.e., non-frustrated) couplers and biases as
    \begin{subequations}
    \label{eq:satisfied}
        \begin{align}
            \label{eq:satisfiedJ}
            \mathcal{J}^{\alpha} &= \{(i,j) \mid J_{ij}\alpha_i\alpha_j<0 \}, \\
            \label{eq:satisfiedH}
            \mathcal{H}^{\alpha} &= \{i \mid h_i\alpha_i<0\}.
        \end{align}
    \end{subequations}
These sets consist of those couplings and biases of \equref{eq:H_EA} that are satisfied in the state $\alpha$, meaning that the corresponding terms in the Hamiltonian contribute to a reduction of the total energy of the system.

Intuitively, the presence of competing spin-spin interactions in \equref{eq:H_EA} causes frustration, which prevents the ground state from establishing a simple long-range (anti-)ferromagnetic order. Instead, the spins align in random directions, forming a glass phase~\cite{SG_Complexity}. This phase is characterized by the existence of exponentially many low-energy local minima, where a \emph{local minimum} refers to a state for which flipping a single (or a few) spins always increases the energy. The energy spectrum of these local minima typically has small gaps, with macroscopic high and wide energy barriers separating them~\cite{SG_Ergodicity}. The latter implies that $\mathcal{O}(N)$ spins would need to be flipped in order to escape a local minimum~\cite{SG_Theory}. Despite the complexity of the energy landscape, the formation of local minima stems from a general structure in the interactions of the spins. Local differences in the spin-spin couplings and biases lead to the formation of spin domains $\mathcal{D}_1, \, \dots \, , \, \mathcal{D}_{N_{\mathrm{dom}}}\subset\{0, \, \dots \, , \, N-1\}$, representing groups of highly correlated spins~\cite{SG_Theory, SG_Domains_Annealingtime}.

\subsection{Spin domains}
We define a \emph{spin domain} $\mathcal{D}_k\subset\{0, \, \dots \, , \, N-1\}$, with $k=1, \, \dots \, , \, N_{\mathrm{dom}}$, as a maximally large connected sub-spin glass of \equref{eq:H_EA}, such that its energy landscape is \emph{ergodic} with respect to spin-reversal transformations. This means that the ground state of the domain is accessible from any state through a sequence of decreasing energies using solely single spin flips and state inversions (i.e.~flipping all spins). In computational terms, \emph{ergodic} means that the global minimum of the domain can be reached via a greedy search algorithm. Note that a domain can consist of a single spin if the local magnetic field $h_i$ dominates the interaction. An example $24$-qubit spin glass with $N_{\mathrm{dom}}=3$ domains is shown in Fig.~\ref{fig:spin_glass_graph}a. The respective energy spectra of the spin domains are given in panels (a.1-3).

Each spin domain $\mathcal{D}_k$ identifies a trivial sub-spin glass, defined by the quadratic terms $i,j\in\mathcal{D}_k$ within the domain, 
    \begin{align}
        \sum_{i,j\in\mathcal{D}_k} J_{ij}\sigma_i^z\sigma_j^z.
    \end{align}
This sub-spin glass has two degenerate ground states $\beta^{*(k)}$ and $\overline{\beta^{*(k)}}$, due to the $\mathbb{Z}_2$ symmetry of the spin-spin couplers, with $\overline{\beta^{*(k)}}$ denoting the inverse state of $\beta^{*(k)}$ (i.e., all spins flipped). By combining the ground states of all domains, the $2^{N_{\mathrm{dom}}}$ \emph{local minima} of the full spin glass can be constructed. Consequently, these local minima form an $N_{\mathrm{dom}}$-dimensional hypercube, with the edges denoting state inversions of single spin domains (see Fig.~\ref{fig:spin_glass_graph}b).

The complexity of the spin-glass energy landscape stems from the domain borders, where spin domains couple to each other (i.e.~spin-spin couplings $J_{ij}$ between domains) and to the local biases $h_i$. These interactions lift the degeneracies of the domains' ground states, assigning energy shifts to the local minima and forming a hierarchy on the hypercube~\cite{SG_Hierarchy}. 

For the three-dimensional hypercube shown in Fig.~\ref{fig:spin_glass_graph}b, Fig.~\ref{fig:spin_glass_graph}c visualizes the energy landscape obtained by unfolding its facets. Note that the hardness of spin glasses arises from the lack of knowledge about the exact domains $\mathcal{D}_k$. If the distribution of spin domains were known, the complexity of the problem would be reduced from $2^N$ to $2^{N_{dom}}$ with the number of domains $N_{dom} \ll N$.

\subsection{Distance measures}
For the following description of LDA, a distance measure between arbitrary spin configurations $\alpha$ and $\beta$ is required. A commonly used metric for this purpose is the Edward-Anderson order parameter~\cite{SG_Theory_EA}, measuring the overlap of the states $\alpha$ and $\beta$ as
    \begin{align}
        q_{EA}(\alpha, \beta) = \frac{1}{N} \sum^{N-1}_{i=0} \alpha_i\beta_i.
        \label{eq:q_EA}
    \end{align}
By definition, $q_{EA}\in[-1,1]$, with $q_{EA} = 1 \Leftrightarrow \alpha=\beta$ and $q_{EA} = -1 \Leftrightarrow \alpha=\overline{\beta}$. Note that $q_{EA}$ is directly related to the Hamming distance $d_{\mathrm{ham}}$ between the bit strings \texttt{a} and \texttt{b} via $d_{\mathrm{ham}} = N(1-q_{EA})/2$.
    
While $q_{EA}$ is a sufficient tool for the statistical analysis of phase transitions in disordered systems~\cite{King2023QuantumCriticalDynamics5000SpinGlass}, it is indifferent to the distance between the energies $E_\alpha$ and $E_\beta$ of the states. This means that two states with the same order parameter $q_{EA}$ can have significantly different energies, depending on the amount of frustration that they cause. To compensate for this, we propose a quantity $q_F(\alpha,\beta)$ that evaluates the similarity between a reference state $\alpha$ and a state $\beta$ based on the sets of satisfied couplers and biases (see~\equref{eq:satisfied}). We define $q_F$ as the fraction of couplers and biases that are simultaneously satisfied by $\alpha$ and $\beta$ (i.e., $\mathcal{J}^{\alpha}\cap\mathcal{J}^{\beta}$ and $\mathcal{H}^{\alpha}\cap\mathcal{H}^{\beta}$) to the total set of satisfied terms by $\alpha$ (i.e., $\mathcal{J}^{\alpha}$ and $\mathcal{H}^{\alpha}$),
    \begin{align}
        q_F(\alpha, \beta) = \frac{\sum\limits_{\substack{(i,j)\in \mathcal{J}^{\alpha}\cap\mathcal{J}^{\beta}\\\mathrm{with}\:(\alpha_i, \alpha_j) = (\beta_i, \beta_j)}} |J_{ij}| + \sum\limits_{i\in \mathcal{H}^{\alpha}\cap\mathcal{H}^{\beta}} |h_i|}{\sum\limits_{(i,j)\in \mathcal{J}^{\alpha}} |J_{ij}| + \sum\limits_{i\in \mathcal{H}^{\alpha}} |h_i|}.
        \label{eq:q_F}
    \end{align}
By definition, $q_F \in [0,1]$, with $q_F = 1\;(0)$ denoting that $\mathcal{H}^{\beta} \subset \mathcal{H}^{\alpha}$ $\left(\mathcal{H}^{\beta} \cap \mathcal{H}^{\alpha} = \emptyset\right)$ and $\mathcal{J}^{\beta} \subset \mathcal{J}^{\alpha}$ $\left(\mathcal{J}^{\beta} \cap \mathcal{J}^{\alpha} = \emptyset\right)$. We remark that in the singular case $\mathcal{H}^{\alpha} = \mathcal{J}^{\alpha} = \emptyset$, we define $q_F = 1$. The additional condition $\alpha_i=\beta_i$ in the first summation of \equref{eq:q_F} ensures that the spins in the simultaneously satisfied couplers $J_{ij}$ are also aligned with one another, $(\alpha_i,\alpha_j) = (\beta_i,\beta_j)$. This is necessary because $J_{ij}\alpha_i\alpha_j=J_{ij}\beta_i\beta_j < 0$ would also be satisfied if $(\alpha_i,\alpha_j) = (\overline{\beta_i},\overline{\beta_j})$. Hence, the condition lifts the spin-reversal symmetry of the spin domains, in order to account for the Hamming distance between the states. 

\begin{figure}
    \centering
    \includegraphics[width=\linewidth]{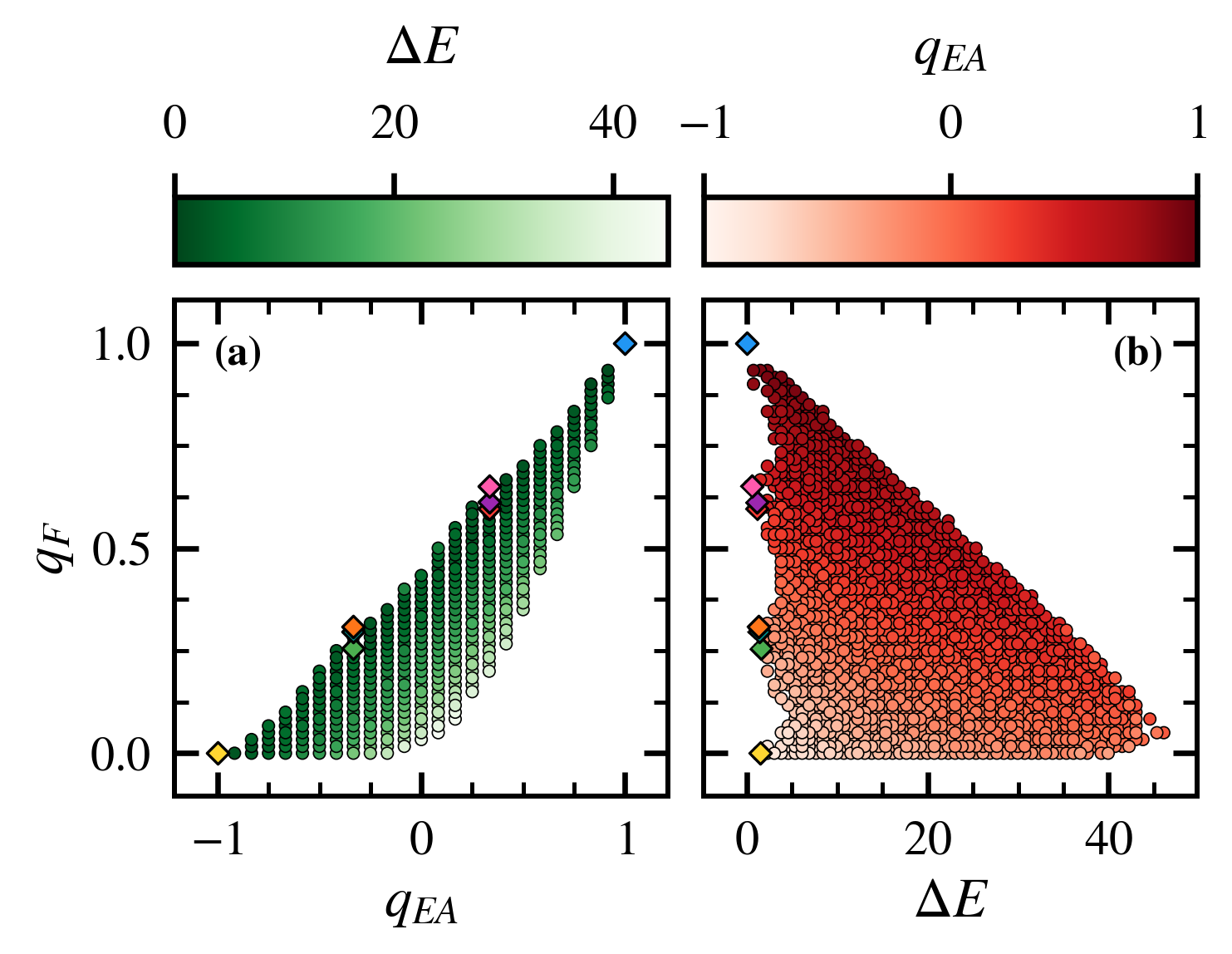}
    \caption{Comparison of the quanitiy $q_F(\alpha^*, \beta)$ to (a) the Edward-Anderson order parameter $q_{EA}(\alpha^*, \beta)$ and (b) the energy difference $\Delta E = E_{\beta} - E_{\alpha^*}$ for states $\beta$ w.r.t.~the global minimum $\alpha^*$ of the $24$-qubit spin-glass shown in Fig.~\ref{fig:spin_glass_graph}a. The colored diamonds mark the eight local minima (see \figref{fig:spin_glass_graph}c).}
  \label{fig:comp_orderparams}
\end{figure}

The definition of $q_F$ is based on the observation that each local minimum $\alpha^*$ can be uniquely identified (up to state inversion in case of $h_i = 0$) by the sets of satisfied couplers $\mathcal{J}^{\alpha^*}$ and biases $\mathcal{H}^{\alpha^*}$ (see Fig.~\ref{fig:spin_glass_graph}c.1-3, where the local minima satisfy different couplers between the domains). By ranking each state according to its energy in the sub spin-glass $\left(\mathcal{J}^{\alpha^*},\,\mathcal{H}^{\alpha^*}\right)$, the parameter $q_F$ integrates both the Hamming distance ($q_{EA}$) and the energy distance ($\Delta E$) from $\alpha^*$ into the similarity measure (see Fig.~\ref{fig:comp_orderparams}a). This allows $q_F$ to distinguish the energy valley of $\alpha^*$ (i.e. the neighborhood of states around $\alpha^*$, where some domains are in an excited state) from the other local minima in Fig.~\ref{fig:spin_glass_graph}c. As a result, only states with both a small Hamming distance and energy distance are considered close to $\alpha^*$, making $q_F$ indifferent to low-energy states of other energy valleys (see peaks at $\Delta E \approx 0$ in Fig.~\ref{fig:comp_orderparams}b). 

A key property of $q_F$ is its asymmetry (i.e. $q_F(\alpha, \beta) \neq q_F(\beta, \alpha)$), since \equref{eq:q_F} considers only terms that are satisfied by the reference state $\alpha$, but ignores additional terms satisfied by $\beta$. As a consequence, a state $\beta$ deeper in the energy valley, with $\mathcal{J}^{\beta} \supset \mathcal{J}^{\alpha}$ and $\mathcal{H}^{\beta} \supset \mathcal{H}^{\alpha}$, typically has a larger $q_F$ value than a higher-energy state, which violates terms in $\mathcal{H}^{\alpha}$ and $\mathcal{J}^{\alpha}$. To make the meaning of $q_F$ more intuitive, an explicit example of this asymmetry for a $16$ spin-glass instance is shown in App.~\ref{app:asymmetry_qf}. This means that $q_F$ has an intrinsic bias towards lower-energy states, which will be crucial for the definition of LDA in Sec.~\ref{sec:lda}.


\section{Adiabatic quantum annealing}
\label{sec:qa}
Adiabatic quantum annealing, as proposed by Farhi et al.~\cite{QA_Theory_2}, is a procedure for solving NP-hard combinatorial optimization problems (COP) through quantum fluctuations. With the search space of the COP encoded in the eigenspectrum of a problem Hamiltonian $H_P$ (e.g.~\equref{eq:H_EA}), the system is complemented by a driving Hamiltonian $H_D$ that is non-diagonal in the eigenbasis of $H_P$. Since the spin-glass Hamiltonian \equref{eq:H_EA} is diagonal in the computational basis, we consider the transverse field Ising model
    \begin{align}
        H_{QA}(t) = H_P + \Gamma(t) \cdot H_D \;\;\text{with}\;\; H_D = - \sum^{N-1}_{i = 0} \sigma^x_i,
        \label{eq:H_QA}
    \end{align}
which is closely related to the design of D-Wave quantum annealers (see App.~\ref{app:qpu_settings}). $\Gamma(t)$ is the annealing schedule, defining the relative strength between $H_P$ and $H_D$ over time $0 \leq t \leq T$. Initially, $\Gamma(t=0) \gg J$, such that the instantaneous ground state (GS) is approximately given by $\ket{+}^{\bigotimes N}$ (where $\ket+=(\ket0+\ket1)/\sqrt2$), which is initially separated by a large energy gap $\propto \Gamma(0)$ from the rest of the spectrum. Operating the system at a temperature that is much smaller than the scale set by $\Gamma(0)$ is supposed to initialize it in the GS of $H_D$. To reach the GS of $H_P$, $\Gamma(t)$ is decreased adiabatically to zero as $t\to T$, allowing the system to follow the instantaneous GS of $H_{QA}(t)$~\cite{QA_Theory_2, QA_Theory_1}. In order to prevent Landau-Zener transitions into excited states during this process, the adiabatic theorem requires $T^{-1} \ll \Delta^2_{min}$~\cite{QA_Adiabatic_Condition}, where $\Delta_{min}$ denotes the minimal energy gap between the two lowest instantaneous eigenstates of $H_{QA}(t)$.

\begin{figure}
    \centering
    \includegraphics[width=\linewidth]{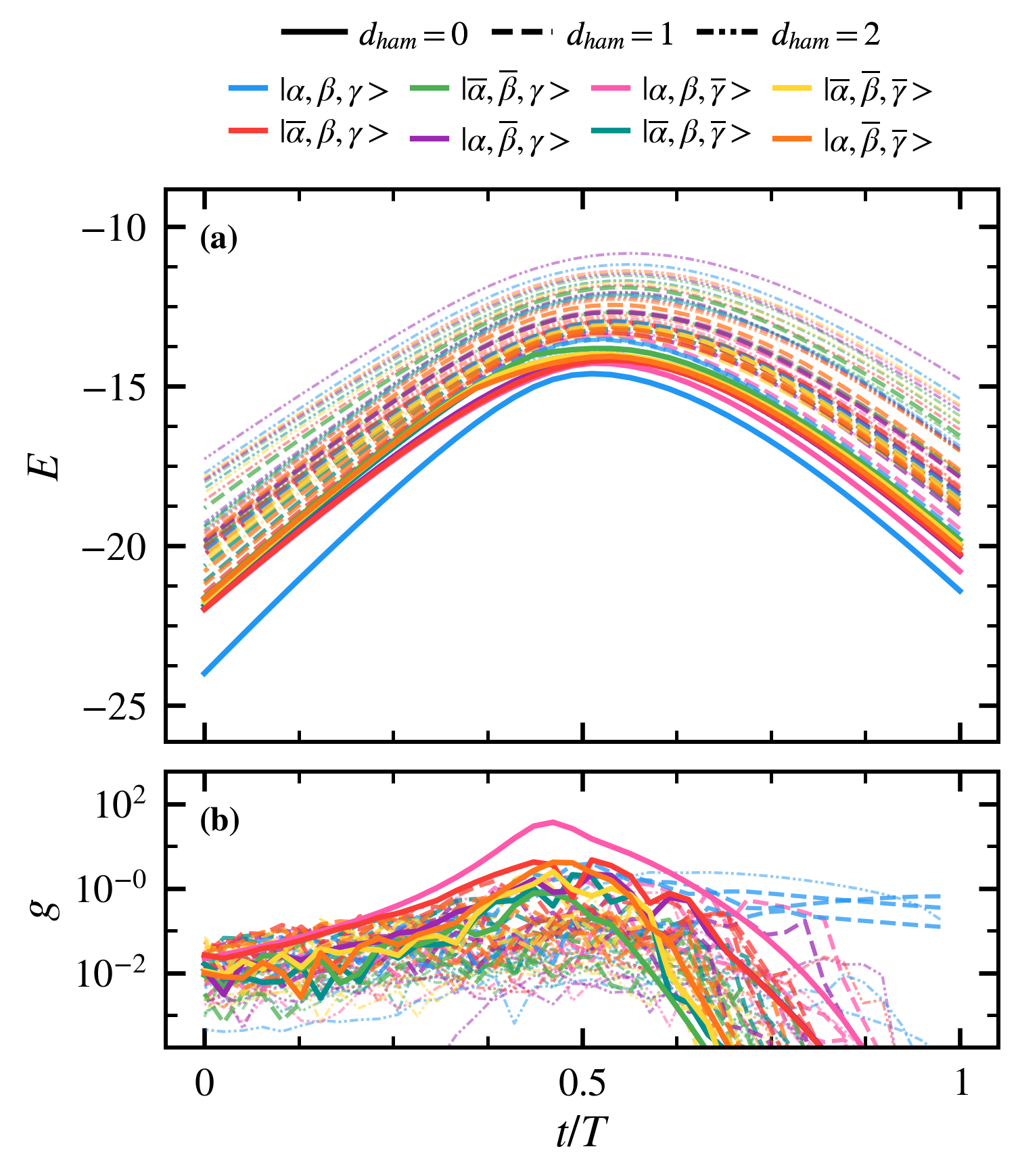}
    \caption{(a) Instantaneous energy spectrum of the transverse field Ising model (see~\equref{eq:H_QA}) for the $24$-qubit spin-glass instance shown in Fig.~\ref{fig:spin_glass_graph}a, using a linear annealing schedule $\Gamma(t)$. The plot presents a selection of energy levels corresponding to the $8$ valleys in the energy landscape (see Fig.~\ref{fig:spin_glass_graph}c). For each valley, the local minimum (solid lines) is shown along with three states at Hamming distance $d_{ham} = 1$ (dashed line) and three states at $d_{ham} = 2$ (dash-dotted lines), representing the lowest-energy states along the edges of the hypercube. (b) Adiabatic ratios (see~\equref{eq:g}) for the states presented in panel (a). The 24-qubit eigenstates and eigenvalues in the vicinity of each local minimum were obtained numerically by solving the time-dependent Schrödinger equation (see e.g.~\cite{TDSE}) for reverse-annealing starting in the given states.}
  \label{fig:energyspectrum_g_orig}
\end{figure}

The minimum annealing time $T$ is the main measure for the computational complexity of adiabatic quantum annealing. Significant evidence exists that the presence of first-order quantum phase transitions (QPTs) during the adiabatic evolution causes an exponentially fast closure of $\Delta_{min}$ with the system size $N$, resulting in exponentially long annealing times~\cite{QA_QPT_1,QA_QPT_2}. The authors in~\cite{SG_Domains_Annealingtime} argue that $\Delta_{min}$ scales inversely proportional to the square root of the number of competing local minima (and states near them) with energies close to the global minimum and a large Hamming distances $d_{\mathrm{ham}}$. Regarding the spin-glass instances under investigation, the number of local minima scales exponentially, $2^{N_\mathrm{dom}}$, with the number of domains $N_\mathrm{dom} \propto N$. 

Figure~\ref{fig:energyspectrum_g_orig}a presents the instantaneous energy spectrum of the $24$-qubit spin-glass instance shown in Fig.~\ref{fig:spin_glass_graph}a, with the adiabatic ratio~\cite{QA_Adiabatic_Ratio} 
    \begin{align}
        g(t) = \frac{\abs{\bra{m(t)} \dot{H}_{QA}(t) \ket{GS(t)}}}{\abs{E_m(t) - E_{GS}(t)}^2}
        \label{eq:g}
    \end{align}
depicted in Fig.~\ref{fig:energyspectrum_g_orig}b for the states marked in Fig.~\ref{fig:spin_glass_graph}c. Here, $\ket{m(t)}$ ($\ket{GS(t)}$) denotes the instantaneous $m$-th excited (ground) state, $\dot{H}_{QA}(t)$ is the $t$ derivative of the instantaneous Hamiltonian and $E_{m}(t)$ ($E_{GS}(t)$) corresponds to the eigenenergies of the $m$-th excited (ground) state.

In agreement with~\cite{QA_QPT_3, QA_QPT_4}, the system undergoes a second-order QPTs at $t_c/T \approx 0.45$, by evolving from the paramagnetic phase (localized in the $\sigma_x$ basis) to an ordered phase (localized in the $\sigma_z$ basis). As a result, the system transitions from a delocalized state, where it forms a superposition across all valleys, to a semi-localized state confined to a single valley~\cite{QA_QPT_1}. During the QPT, all valleys are coupled, as indicated by the peaks in $g$ for the $8$ global and local minima (see solid lines in Fig.~\ref{fig:energyspectrum_g_orig}b). Additionally, the instantaneous energy spectrum exhibits transitions between states within the same valley. Since these states differ by small Hamming distances, they remain coupled after the second-order QPTs for $t > t_c$ (see blue dashed and dashed-dotted curves in Fig.~\ref{fig:energyspectrum_g_orig}b). 

While the second-order QPT is generally associated with a polynomial scaling of $T$, the presence of additional anti-crossings at $t > t_c$ between semi-localized states in large spin-glass instances induces first-order QPTs~\cite{QA_QPT_1, QA_QPT_4}. These transitions correspond to quantum tunneling events, where the system relocates entirely from one valley to another, leading to energy gaps $\Delta$ that close exponentially with the Hamming distance $d_{ham}$ between valleys. Since $d_{ham}$ scales linearly with the system size $N$ in our spin-glass instances, this results in an exponential scaling of $T$~\cite{QA_QPT_2, QA_QPT_3}. Recently, locally optimized annealing procedures have been proposed, which reduce $T$ by decelerating the annealing schedule $\Gamma(t)$ at the vicinity of the QPT~\cite{QA_Local_Optimized_1, QA_Local_Optimized_2}. However, these strategies can only provide a quadratic speed up~\cite{SG_Domains_Annealingtime}. Since the required annealing times still exceed the accessible range on real devices, it is in practice impossible to solve large spin glasses using a single annealing run. 

To overcome this bottleneck, several authors proposed the use of iterative protocols to incorporate classical information about the ground state into the annealing process~\cite{Cyclic_annealing_1, Ra_Hg, IRA_Theory_3}. The reverse annealing protocol, as proposed by Chancellor~\cite{IRA_Theory_1} and Yamashiro et al.~\cite{IRA_Theory_2}, has been implemented on D-Wave QPUs. It starts by initializing the system in a computational basis state (e.g.~obtained by a previous anneal or a classical heuristic), with $\Gamma=0$. Quantum fluctuations are slowly introduced by increasing $\Gamma$ to an intermediate value $\Tilde{\Gamma}$ at $t = T_s$, where the process is paused for a period $\tau$ until $t = T_e$. The protocol finishes by annealing back to $\Gamma=0$ and measuring the qubits. By repeating this cycle and initializing each reverse anneal by the measurement of the previous, iterative reverse annealing (IRA) is hoped to move the system closer to the true ground state and overcome the above-mentioned limitation of an exponential annealing time $T$.

In practice, however, reverse annealing performs a dissipative local search into the neighborhood of the initial state~\cite{IRA_Failing_1}. The spread of this search is controlled by the inversion point $\Tilde{\Gamma}$ and the QPTs it traverses. If no QPT is crossed, the system spreads locally (in Hamming distance) within its energy valley. However, if $\Tilde{\Gamma}$ is chosen large enough, such that a QPT is crossed, the system explores the search space globally by tunneling into other energy valleys~\cite{IRA_Theory_1}. Note that $T$ must be sufficiently short, such that the process is non-adiabatic, because otherwise the system would simply return to its initial state.

In a closed system, there is no direct mechanism in the reverse annealing protocol to favor the search of lower-energy states~\cite{IRA_Failing_1}. Instead, the system spreads uniformly in energy around the initial state, gravitating (over multiple iterations) towards an equilibrium state, where the number of states with lower and higher energies balance. This makes it exponentially hard to find the ground state for $N\gg1$. 

On real devices, however, the QPU is always in interaction with its surrounding environment. In such open systems, thermal relaxation moderates the transition between neighboring states~\cite{IRA_Thermal_1, IRA_Thermal_2}. At sufficiently low temperatures, this causes a repopulation of lower-energy states during the pause, which allows reverse annealing to improve on the initial state. Passarelli~et.~al.~\cite{IRA_Failing_2} argue that thermal relaxation is the main mechanism driving the performance of reverse annealing. This means that the success of IRA depends mainly on the effective temperature of the QPU. This gives an intuition as to why the D-Wave Advantage1 $5.4$~QPU operating at $\approx16.4mK$~\cite{QPU_Advantage_5_4} is not able to find the ground state of the large spin glasses under investigation through IRA (see \secref{sec:benchmarks}).


\section{Learning-Driven Annealing}
\label{sec:lda}
With Learning-Driven Annealing (LDA) we introduce a framework for linking individual QA runs into a global solution strategy to mitigate hardware limitations, such as finite annealing times and integrated control errors. LDA differs from other iterative strategies in that it does not tune the annealing procedure (e.g.~annealing time or annealing schedule), but instead learns about the problem structure (i.e.~spin domains) to systematically \emph{modify} the problem Hamiltonian. LDA acts on the instantaneous energy spectrum in order to change the strength of QPTs between the energy valleys. This focuses the annealing evolution into a low-energy region of the Hilbert space by suppressing QPTs into high-energy valleys. LDA achieves this by replacing the problem Hamiltonian with a so-called \emph{feature Hamiltonian} in the forward anneal. The construction of this Hamiltonian is based on the definition of the quantity $q_F$ (see~\equref{eq:q_F}) and adapts the strength of biases and couplers in $H_P$ based on learned information about the domain structure to energetically isolate low-energy valleys in the instantaneous energy spectrum. This allows the annealing evolution to successively reach deeper states in the energy landscape.

\begin{figure*}
  \centering
  \includegraphics[width=\linewidth]{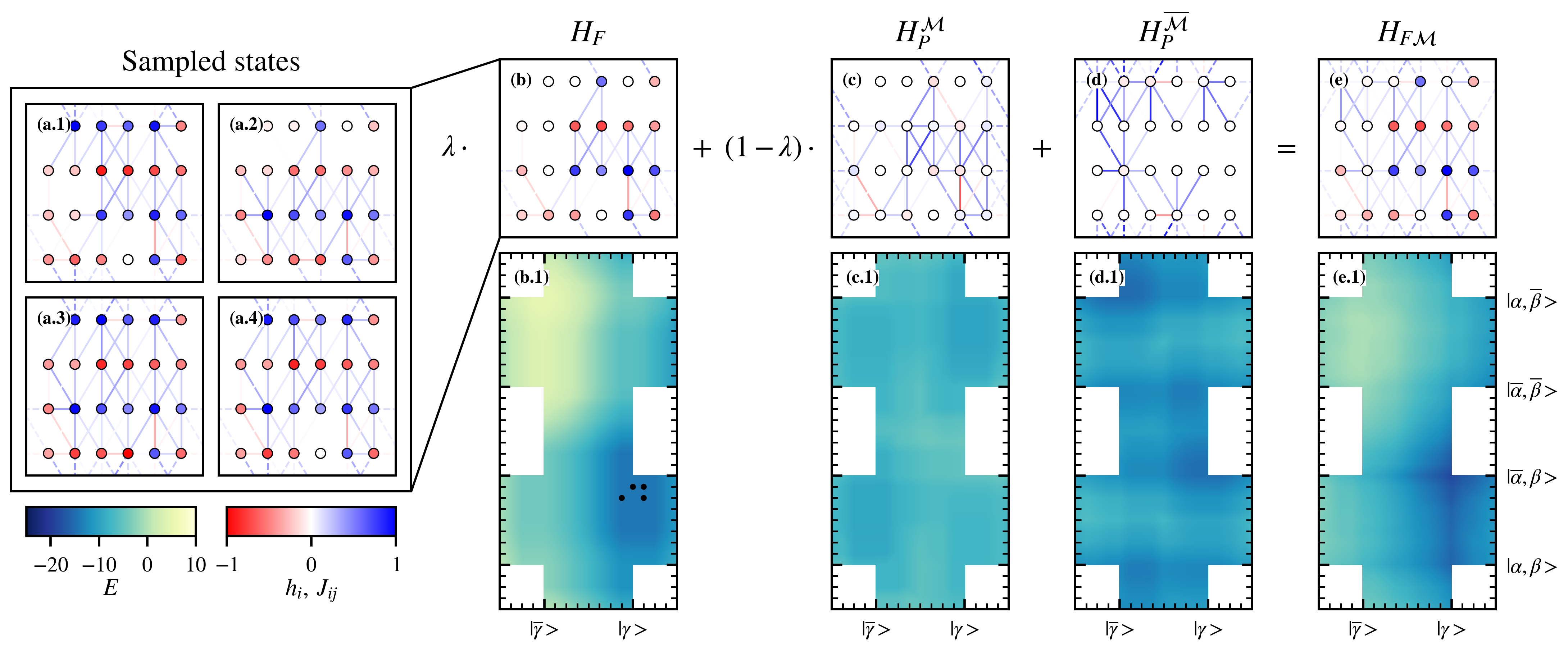}
  \caption{Construction of $H_{F\mathcal{M}}$ (e, \equref{eq:H_F_M}) for the $24$-qubit spin-glass instance shown in Fig.~\ref{fig:spin_glass_graph}, with $\lambda = 0.1$. The feature Hamiltonian $H_F$ (b) is derived from four sampled states (a.1-4) using \equref{eq:H_F_mod} and the bitmask $\mathcal{M}$, with $m_i = 1$ if the sampled states agree in the $i$-th bit (this is the case for all non-white nodes shown in panel (b)). Panels (c) and (d) depict subsets of the original problem Hamiltonian, constructed using \equref{eq:H_P_mod}, with (c) applying to the terms included in $H_F$, and (d) addressing the remaining biases and couplers. Panels (b-e.1) illustrate the corresponding energy landscapes of the Hamiltonians in (b-e), with the four sampled states shown as black dots in (b.1).}
  \label{fig:lda}
\end{figure*}

The feature Hamiltonian, denoted as $H_{F}(\alpha)$, is defined w.r.t.~a classical state $\alpha = \alpha_{N-1} \cdots \alpha_1 \alpha_0$ for a spin glass instance $P = \left(h, J\right)$ as
    \begin{align}
        H_{F}(\alpha) &= \sum\limits_{(i,j)\in \mathcal{J}^{\alpha}} K_{ij} + \sum\limits_{i\in \mathcal{H}^{\alpha}} h_i \sigma^z_i 
        \label{eq:H_F}\\
        K_{ij} &= -\frac{\abs{J_{ij}}}{2} \left[\alpha_i \alpha_j \sigma^z_i \sigma^z_j + \alpha_i \sigma^z_i + \alpha_j \sigma^z_j \right].
        \label{eq:H_F_K}
    \end{align}
$\mathcal{H}^{\alpha}$ and $\mathcal{J}^{\alpha}$ denote the sets of satisfied biases and couplers, respectively (see~\equref{eq:satisfiedH} and~\eqref{eq:satisfiedJ}). $H_{F}(\alpha)$ transforms the energy spectrum of $P$ by retaining only terms that are satisfied by $\alpha$ in $P$. This \emph{deforms} the energy landscape according to the quantity $q_F$ (see \equref{eq:q_F}), by arranging the states based on both their Hamming distance and energy distance from $\alpha$, with $\alpha$ becoming the new ground state (see Thm.~\ref{lem:gs}).

In \equref{eq:H_F} we use the transformation $J_{ij}\sigma_i^z\sigma_j^z \xrightarrow{} K_{ij}$ to lift the $\mathbb{Z}_2$ symmetry of the spin-spin couplers $J_{ij}$ by introducing a penalty if either of the two spins $i$ and $j$ does not match the reference state $\alpha$, i.e.,
    \begin{align}
        K_{ij}\ket{\alpha_i, \alpha_j} &= -\frac{3}{2} \abs{J_{ij}}, \qquad K_{ij}\ket{\alpha_i, \overline{\alpha_j}} = \frac{1}{2} \abs{J_{ij}}\\
        K_{ij}\ket{\overline{\alpha_i}, \alpha_j} &= \frac{1}{2} \abs{J_{ij}}, \qquad\;\;\; K_{ij}\ket{\overline{\alpha_i}, \overline{\alpha_j}} = \frac{1}{2} \abs{J_{ij}}.
    \end{align}
Note that a factor of $1/2$ is introduced in \equref{eq:H_F_K} to ensure that the weighting of the modified couplers remains consistent with the original problem Hamiltonian, with energy gaps of $2\abs{J_{ij}}$. This transformation corresponds to the condition $\alpha_i = \beta_i$ in \equref{eq:q_F} and removes the degeneracy of the spin domains (i.e., each domain now has only one GS). Since the domain's degeneracy is responsible for the existence of energy valleys (see Fig.~\ref{fig:spin_glass_graph}c), removing the $\mathbb{Z}_2$ symmetry flattens the energy landscape, while the restriction of $H_{F}(\alpha)$ to $\mathcal{H}^{\alpha}$ and $\mathcal{J}^{\alpha}$ deforms the landscape in the direction of the reference state $\alpha$ (see Fig.~\ref{fig:lda}b.1). As a result, $H_{F}(\alpha)$ is frustration-free (i.e., all couplers and biases are satisfied in the GS $\alpha$) and the modified energy landscape now increases monotonically with $q_F$ from $\alpha$ (see Fig.~\ref{fig:comp_orderparams}). We remark that the monotonicity holds only for the feature Hamiltonian $H_F$, which will be used henceforth to construct the updated problem Hamiltonian (see Eq.~\ref{eq:H_F_M} below).

Since the similarity measure $q_F$ is asymmetric, a state $\beta$ that lies deeper in the energy valley than $\alpha$ (i.e., it is closer to the local minimum $\alpha^*$ in terms of $q_F$) satisfies most of the biases and couplers already satisfied by $\alpha$, along with additional terms. Consequently, $\mathcal{H}^{\alpha} \subsetsim \mathcal{H}^{\beta}$ and $\mathcal{J}^{\alpha} \subsetsim \mathcal{J}^{\beta}$, such that $q_F(\alpha, \beta) \approx 1$. In this case, $\beta$ is either a degenerate ground state or a weakly excited state of the feature Hamiltonian $H_{F}(\alpha)$. In contrast, a higher-energy state in $P$ necessarily frustrates terms satisfied by $\alpha$, making it a highly excited state of $H_F(\alpha)$ (cf.~App.~\ref{app:asymmetry_qf} for a 16-qubit example). As a result, the modified energy landscape is nearly flat in the region of Hilbert space lying deeper in the energy valley than $\alpha$, termed the \emph{search region}, while outside this region the energy increases monotonically with the Hamming distance from $\alpha$ (see Thm.~\ref{lem:monotonic} in App.~\ref{app:prop_Hf}). The size of the search region depends on the elements of $\mathcal{H}^{\alpha}$ and $\mathcal{J}^{\alpha}$, referred to as \emph{features}. These features represent the information provided by $\alpha$ about the location of a nearby local minimum $\alpha^*$. If $\alpha$ is a high-energy state, most biases and couplers in $P$ are frustrated, leading to a highly degenerate $H_{F}(\alpha)$ with a large search region, as $\mathcal{H}^{\alpha}$ and $\mathcal{J}^{\alpha}$ contain only few elements. This situation changes as $\alpha$ approaches a local minimum $\alpha^*$. As the overlap between the satisfied terms of $\alpha$ and $\alpha^*$ increases, $H_{F}(\alpha)$ incorporates more features characterizing the local minimum, resulting in a progressively deformed energy landscape that narrows the search region around $\alpha^*$.

The monotonicity of the modified energy landscape outside the search region suppresses all avoided level crossings in the original instantaneous spectrum that would drive the system out of the search region, as they are now separated by large energy gaps. In particular, second-order QPTs that induce delocalized-to-localized transitions into competing valleys are removed, since these valleys are flattened in the modified landscape. Thus, the feature Hamiltonian $H_F(\alpha)$ energetically isolates an annealing path (i.e., a sequence of transitions and non-transitions at avoided level crossings) in the original instantaneous spectrum that confines the system to the search region. The length of this annealing path (i.e., the number of transitions fixed by $H_F(\alpha)$) depends on the features provided by $\alpha$ and is therefore related to the size of the search region.

In a forward anneal, the feature Hamiltonian confines the dynamics to the search region. However, because the energy landscape remains nearly flat within this region, it cannot resolve the original valley structure and identify the minimum $\alpha^*$. As a result, the system evolves into an almost equal superposition over the search space. To address this, we reintroduce the original problem Hamiltonian $H_P$ within the search region, yielding:
\begin{align}
        H_{F}(\alpha, \mathcal{M}) &= \sum\limits_{\substack{(i,j)\in \mathcal{J}^{\alpha}\\\mathrm{with}\:m_i=m_j=1}} \lambda \cdot K_{ij} + \sum\limits_{\substack{i\in \mathcal{H}^{\alpha}\\\mathrm{with}\:m_i=1}} \lambda \cdot h_i \sigma^z_i,
        \label{eq:H_F_mod}\\
        H_{P}(\mathcal{M}) &= \quad \sum\limits_{i<j} \begin{cases}
                                        \left(1 - \lambda\right) \cdot J_{ij} \sigma^z_i \sigma^z_j, & \text{if } m_i, m_j = 1\\
                                        J_{ij} \sigma^z_i \sigma^z_j, & \text{else}
                                    \end{cases}\\
                & \quad + \sum\limits_{i}   \begin{cases}
                                        \left(1 - \lambda\right) \cdot h_{i} \sigma^z_i, & \text{if } m_i = 1\\
                                        h_{i} \sigma^z_i, & \text{else}
                                    \end{cases}
        \label{eq:H_P_mod}\\
        H_{F\mathcal{M}}(\alpha, \mathcal{M}) &= H_{F}(\alpha, \mathcal{M}) + H_{P}(\mathcal{M})
        \label{eq:H_F_M}
    \end{align}
with the bitmask $\mathcal{M} = m_{N-1} \, \dots \, m_1 \, m_0$ controlling the mixing of $H_P$ and $H_{F}(\alpha)$ and the mixing parameter $\lambda \in [0,1]$ specifying the strength of the deformation of the energy landscape introduced by $H_F$. Note that for $\lambda = 0$, one recovers $H_{F\mathcal{M}}(\alpha, \mathcal{M}) = H_P$, independent of $\mathcal{M}$. $H_{F\mathcal{M}}$ restores the original energy relationship between the states in a local subspace of the hypercube defined by $\mathcal{M}$, while maintaining the deformed energy landscape outside. In particular, $\mathcal{M} = 0\dots0$ corresponds to using only the original problem Hamiltonian $H_P$, whereas $\mathcal{M} = 1\dots1$ indicates that only the new feature Hamiltonian $H_F$ is used.

If we construct $\mathcal{M}$ from a set of sampled states belonging to the same valley as $\alpha$ (i.e., setting $m_i = 1$ if all sampled states share the same spin orientation at site $i$, and $m_i = 0$ otherwise), then $H_P$ is reintroduced only within the search region, thus recovering the minimum $\alpha^*$. This is because, spins aligned across all states of a valley define the subspace of the hypercube in which the valley resides. Hence applying the feature Hamiltonian to these spins removes all competing valleys from the energy landscape. Note that $\alpha$ itself is not necessarily a GS of $H_{F\mathcal{M}}$. In the instantaneous spectrum, this construction restores the avoided crossings and associated dynamics within the search region, while preserving the annealing path that first brings the system into this region. Consequently, LDA enables the system to evolve to a specific point in the original instantaneous spectrum and then resume the dynamics from that point onward. This, in turn, makes it possible to retry difficult segments of the spectrum (i.e., avoided crossings with narrow gaps) until the transition is successfully traversed.

Figure~\ref{fig:lda} illustrates an example of $H_{F\mathcal{M}}$ constructed from four sampled states (black dots in Fig.~\ref{fig:lda}b.1). The bitmask $\mathcal{M}$ has a $1$ at each bit positions where the sampled states agree. The corresponding feature Hamiltonian $H_F$ is shown in Fig.~\ref{fig:lda}b, where the rough energy landscape of the $24$-qubit spin glass (Fig.~\ref{fig:spin_glass_graph}c) is transformed into a more monotonic landscape (Fig.~\ref{fig:lda}b.1). Since the sampled states primarily disagree on the features of the first domain, the landscape remains nearly flat between $\alpha$ and $\overline{\alpha}$. Incorporating the original problem Hamiltonian $H_P$ in this domain (Fig.~\ref{fig:lda}d) allows $H_{F\mathcal{M}}$ resolving the locations of the local minima $\ket{\alpha, \beta, \gamma}$ and $\ket{\overline{\alpha}, \beta, \gamma}$. As a result, all avoided crossings in the instantaneous spectrum are removed, except for the transition between these two valleys. This focuses the annealing dynamics on the relevant subspace (Fig.~\ref{fig:lda}e.1), enabling repeated attempts to traverse the critical transition.


\section{Hybrid optimization}
\label{sec:hybrid_optimization}
LDA gives control over the exploration of the Hilbert space in quantum annealing by integrating learned information about the domain structure into the design of the problem Hamiltonian. By modifying the instantaneous energy spectrum, LDA energetically isolates local minima and creates a more monotonic energy landscape. Consequently, QPTs can be selectively suppressed, restricting the quantum evolution to a subspace of the hypercube.

In this section, we demonstrate how LDA can be used in advanced protocols to solve hard spin-glass instances. We introduce two algorithms, the local search and the global search protocol, which use an iterative application of LDA. Our hybrid optimizer is constructed as an alternating series of these two algorithms.

\subsection{Local search protocol}
\label{subsec:local_search}

The local search protocol is designed to converge from an initial state $\alpha_0$ to a nearby local minimum $\alpha^*$ through an iterative application of LDA. The protocol consists of two phases, a distribution and a convergence phase. 

In the distribution phase, the protocol generates states that spread evenly around the initial state $\alpha_0$ (possibly with higher energies), by biasing the evolution only weakly (i.e.~$\lambda \ll 1$). The goal is to identify those features from $\alpha_0$ that characterize $\alpha^*$, while ignoring terms that are violated in the spin domain ground states. The assumption is that the common features among the sampled states define a subset of $\mathcal{H}^{\alpha^*}$ and $\mathcal{J}^{\alpha^*}$, providing a search area in the energy landscape where the protocol assumes $\alpha^*$ to be located in. In the second phase, the protocol then converges towards $\alpha^*$, by focussing the annealing evolution onto the search area. This is achieved by deforming the energy landscape outside the area using $H_F$ (i.e.~$\lambda \approx 1$). Over multiple iterations of LDA, the sets of sampled states gradually reveal the elements of $\mathcal{H}^{\alpha^*}$ and $\mathcal{J}^{\alpha^*}$, shrinking the search area until the local minimum is found. 

\begin{figure}
  \centering
  \includegraphics[width=\linewidth]{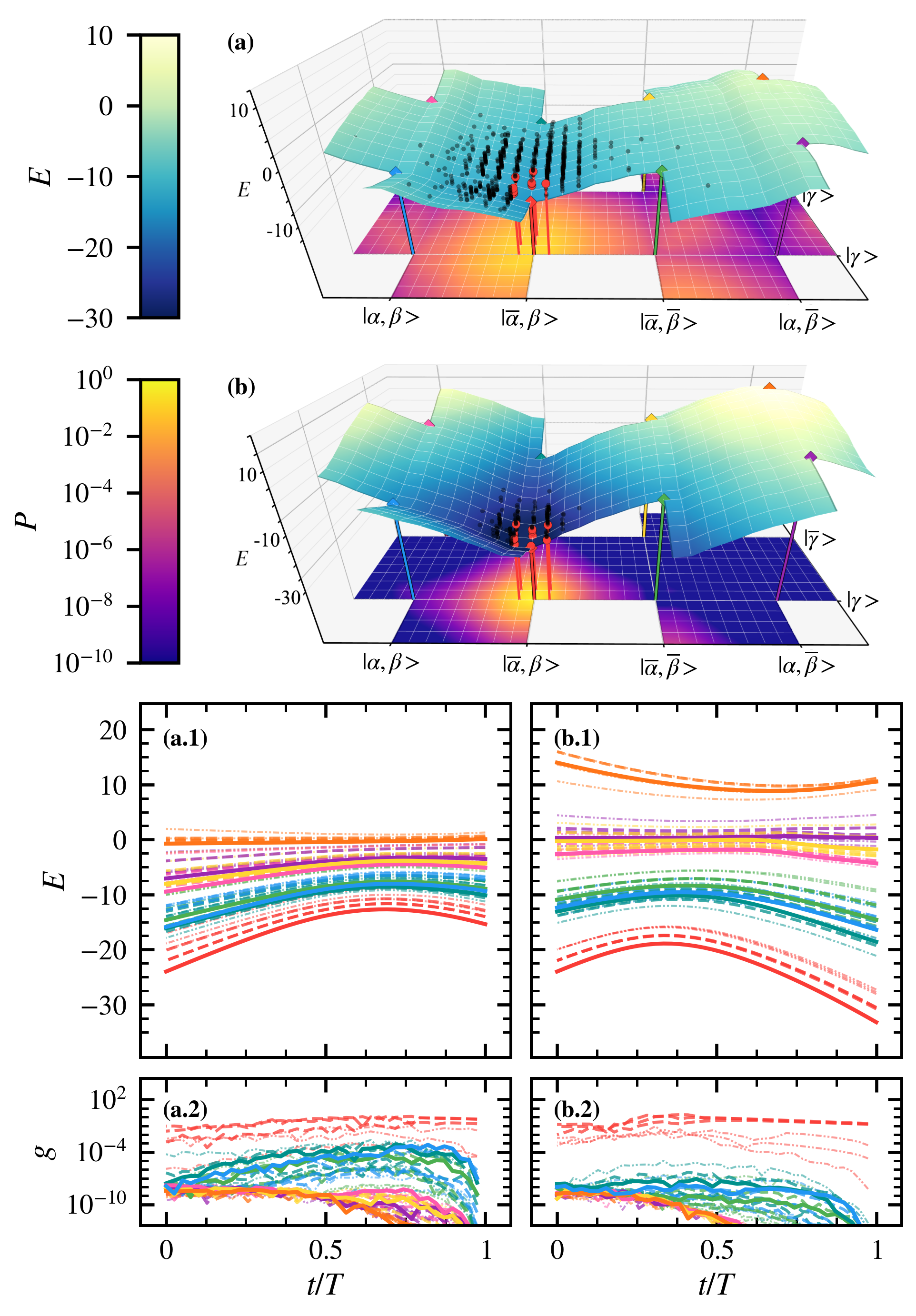}
  \caption{Two iterations of the local search protocol applied to the $24$-qubit spin-glass instance shown in Fig.~\ref{fig:spin_glass_graph}a. The protocol is executed with $Q_{\mathcal{T}} = 0.98$ and $N_{\mathcal{T}} = 5$, using (a) $\lambda = 0.2$ for the first and (b) $\lambda = 1.0$ for the second iteration. The top panels depict the energy landscape ($E$) of $H_{F\mathcal{M}}$ in the first and second iteration, respectively. Black dots represent $2000$ samples from a quantum annealing simulation ($T = 10$), with the probability density ($P$) shown on the bottom plane. Red dots indicate the subset $\mathcal{T}$. The initial state $\alpha_0$ of the first iteration (panel a) is marked as a red dot in Fig.~\ref{fig:spin_glass_graph}c. The color scales are consistent with those in Fig.~\ref{fig:spin_glass_graph}c. Panels (a-b.1) display the instantaneous energy spectrum of the transverse field Ising model with $H_{F\mathcal{M}}$ for each iteration. The energy levels correspond to those in Fig.~\ref{fig:energyspectrum_g_orig}, where the color of the lines indicates the association to the eight valleys in the energy landscape of $H_P$. A linear annealing schedule $\Gamma(t)$ is used throughout as an example. Panels (a-b.2) show the adiabatic ratios $g$ (see \equref{eq:g}) for the states shown in panels (a-b.1), respectively.}
  \label{fig:local_search}
\end{figure}

Each iteration $i$ of the protocol begins with the construction of $H_{F\mathcal{M}}(\alpha_{i}, \mathcal{M}_{i})$ (see~\equref{eq:H_F_M}) using the lowest-energy state $\alpha_{i}$ and the bitmask $\mathcal{M}_{i}$ derived from the previous iteration. For the first iteration ($i=0$, distribution phase), $\mathcal{M}_0 = 1 \dots 1 1$ is used. The hyperparameter $\lambda$ controls the amplification of extracted features in $H_{F\mathcal{M}}$. It functions as an inverse temperature, reflecting the confidence that the extracted features form a subset of $\mathcal{H}^{\alpha^*}$ and $\mathcal{J}^{\alpha^*}$. The goal is to maintain $\alpha^*$ as a ground state of $H_{F\mathcal{M}}$ throughout the iterations. Typically, a smaller $\lambda$ indicates a wider spread of sampled states, as the energy landscape is less deformed (see \figref{fig:local_search}a). We use a geometric schedule 
    \begin{align}
        \lambda(i) = \lambda_s \cdot \left(\frac{\lambda_f}{\lambda_s}\right)^{\frac{i}{I - 1}},
        \label{eq:lambda_i}
    \end{align}
that scales between $\lambda_s$ and $\lambda_f$, with $I$ denoting the total number of Iterations.

In total, $N_{\mathcal{S}}$ states are generated in each iteration and stored in the set $\mathcal{S}$. To construct the bitmask $\mathcal{M}$, the states are sorted in increasing order of their energy in the original problem Hamiltonian $H_P$. A loop then traverses the states $\gamma \in \mathcal{S}$ and includes them in a subset $\mathcal{T}$ if 
    \begin{align}
        \abs{\mathcal{T}} \leq N_{\mathcal{T}} - 1 \quad \text{and} \quad q_F(\beta, \gamma) \leq Q_{\mathcal{T}} \; \forall \; \beta \in \mathcal{T},
        \label{eq:Tau}
    \end{align}
where $N_{\mathcal{T}} < N_{\mathcal{S}}$ and $Q_{\mathcal{T}}$ are LDA parameters. This extracts a dispersed set from the sampled states, where the second condition ensures that all states have a minimum distance w.r.t.~$q_F$ (i.e., they do not cluster together) and are uniformly distributed around $\alpha^*$. The bitmask $\mathcal{M} = m_{N-1} \, \dots \, m_1 \, m_0$ is determined by
    \begin{align}
        m_i =   \begin{cases}
                    1, & \text{if } \abs{\sum_{\beta \in \mathcal{T}} \beta_i} = N_{\mathcal{T}}\\
                    0, & \text{else}
                \end{cases},
        \label{eq:M}
    \end{align}
such that bit $m_i$ is only set if all states $\beta\in\mathcal{T}$ have the same bit $\beta_i$, and $\alpha_{i+1} = \mathcal{S}[0]$. The protocol continues iterating until $\alpha_{i} = \alpha_{i+1}$ or $i = I-1$, with $\alpha_i$ being the local minimum. The full algorithm is detailed in Protocol \ref{algo:local_search}.

Figure~\ref{fig:local_search} shows two iterations of the local search protocol applied to the $24$-qubit spin-glass instance depicted in Fig.~\ref{fig:spin_glass_graph}a, using $Q_{\mathcal{T}} = 0.98$ and $N_{\mathcal{T}} = 5$. The initial state $\alpha_1$ is marked as a red dot in Fig.~\ref{fig:spin_glass_graph}c. Figures~\ref{fig:local_search}a-b.1 demonstrate how the local search progressively identifies the red curves as corresponding to the valley of the local minimum $\left(\overline{\alpha}, \beta, \gamma\right)$, separating it in the instantaneous energy spectrum and suppressing QPTs into other valleys (shown in other colors). Consequently, the evolution is increasingly confined around the local minimum, until it is sampled with high probability.

\begin{algorithm}
    \caption{Local search}
    \label{algo:local_search}
    \begin{algorithmic}[1]
        \vskip 6pt
        
        \Require {
            \begin{minipage}[t]{0.85\linewidth}
                $\alpha_0$ \Comment{Initial state}\\
                $H_P$ \Comment{Problem Hamiltonian}\\
                $N_{\mathcal{S}}$ \Comment{Size of $\mathcal{S}$}\\
                $N_{\mathcal{T}}$ \Comment{Size of $\mathcal{T}$}\\
                $Q_{\mathcal{T}}$ \Comment{Max. similarity between states in $\mathcal{T}$}\\
                $\lambda_s, \lambda_f$ \Comment{Initial and final mixing strength}\\
                $I$ \Comment{Number of iterations}\\
                $T$ \Comment{Annealing time}
            \end{minipage}
        }\vskip 12pt

        \Ensure {
             \begin{minipage}[t]{0.82\linewidth}
                $\alpha_*$ \Comment{Local minimum near $\alpha_0$}
            \end{minipage}
        }\vskip 12pt

        \Procedure{Local search}{}\vskip 2pt

            \State $\mathcal{M} \gets 1 \dots 1 1$\vskip 2pt

            \For{$i$ \texttt{in range} (0, I)}\vskip 2pt
            
                \State $\lambda \gets \texttt{update\_}\lambda\left(i\right)$ \Comment{Eq.~\ref{eq:lambda_i}}\vskip 2pt
        
                \State $H_{F\mathcal{M}} \gets \texttt{generate\_}H_{F\mathcal{M}}\left(H_P, \alpha_{i}, \mathcal{M}, \lambda\right)$\Comment{Eq.~\ref{eq:H_F_M}}\vskip 2pt
                
                \State $\mathcal{S} \gets \texttt{sample}\left(H_{F\mathcal{M}}, N_{\mathcal{S}}, T\right)$\vskip 2pt
                
                \State $\mathcal{T} \gets \texttt{select\_samples}\left(\mathcal{S}, H_P, N_{\mathcal{T}}, Q_{\mathcal{T}}\right)$\Comment{Eq.~\ref{eq:Tau}}\vskip 2pt

                \State $\mathcal{M} \gets \texttt{update\_}\mathcal{M}\left(\mathcal{T}\right)$ \Comment{Eq.~\ref{eq:M}}\vskip 2pt

                \State $\alpha_{i + 1} \gets \mathcal{T}[0]$\vskip 2pt

                \If{$\alpha_{i} = \alpha_{i+1}$}\vskip 2pt
                    \State \Return $\alpha_i$\vskip 2pt
                \EndIf
                
            \EndFor

            \State \Return $\alpha_I$\vskip 6pt
        
        \EndProcedure

    \end{algorithmic}
\end{algorithm}


\subsection{Global search protocol}
\label{subsec:global_search}
The global search protocol is designed to transition from an initial local minimum $\alpha^*$ to a state of a lower-energy valley. The protocol differs from the local search, as it does not directly search around $\alpha^*$, but uses this state as a reference for LDA to gradually filter out higher-energy valleys from the energy landscape. This allows the protocol to overcome large energy barriers, finding lower-energy valleys even at large Hamming distances. Note that the protocol often identifies excited states of the new valley, requiring a subsequent local search to locate the corresponding local minimum.

To reach a new valley in the energy landscape, the global search transitions between the degenerate ground states of at least one domain, by flipping its $\mathcal{O}(N/N_{dom})$ spins in $\alpha^*$. Notably, the procedure does not require knowledge about the domain locations, as the domain structure is inferred from the hierarchy of local minima~\cite{SG_Hierarchy}. This hierarchy arises from differences in domain sizes and inter-domain couplings, leading to different energy gaps between domain ground states. This means that a set $\mathcal{S}$ of energetically similar sampled states (distributed across many valleys) is more likely to align in domains where the two ground states are energetically well-separated, than in domains that only weakly influence the state energies. In such cases, the optimal ground state in these domains can be partially derived from the aligned spins in $\mathcal{S}$. Consequently, domains where the spins in $\mathcal{S}$ and $\alpha^*$ match indicate that $\alpha^*$ resides in the lower-energy ground state. The bitmask $\mathcal{M}$ is then applied to focus the annealing on the remaining domains. Through successive iterations, this strategy gradually substitutes the problem Hamiltonian $H_P$ with the feature Hamiltonian $H_{F}(\alpha^*)$, aligning the domains with $\alpha^*$ and eliminating QPTs into higher energy valleys. Eventually, $H_P$ is only applied to the subspace of suboptimal domains, such that a state from a lower-energy valley can be sampled with high probability.

The protocol begins with the unmodified problem Hamiltonian $H_{F\mathcal{M}} = H_P$ and $\mathcal{M} = 0 \, \dots \, 0 \, 0$. Running LDA $N_{\mathcal{S}}$ times generates the set of sampled states $\mathcal{S}$, from which the subset $\mathcal{T}$ is constructed analogously to the local search protocol (see Eq.~\ref{eq:Tau} in Sec.~\ref{subsec:local_search}). Importantly, $\alpha^*$ is appended to $\mathcal{T}$, and the bitmask $\mathcal{M}$ for the next iteration is determined using Eq.~\eqref{eq:M}. The protocol continues until a state with lower energy than $\alpha^*$ is identified or $i = I - 1$. Notably, $\lambda \approx 1$ is used for fast convergence. The complete algorithm is presented in Protocol~\ref{algo:global_search}.

It is worth noting that the protocol may return to the original valley of $\alpha^*$. To mitigate this, we impose the condition $q_F(\alpha^*, \beta) \leq Q_{\alpha^*} \; \forall \; \beta \in \mathcal{T}$, where the hyperparameter $Q_{\alpha^*}$ denotes the maximum similarity between the selected states and $\alpha^*$, effectively ensuring a minimal Hamming distance between them. This acts as a search radius, with smaller $Q_{\alpha^*}$ values generally leading to valleys farther from $\alpha^*$. This condition can also be applied to previously identified local minima to guarantee the exploration of an unvisited valley.

\begin{figure}
  \centering
  \includegraphics[width=\linewidth]{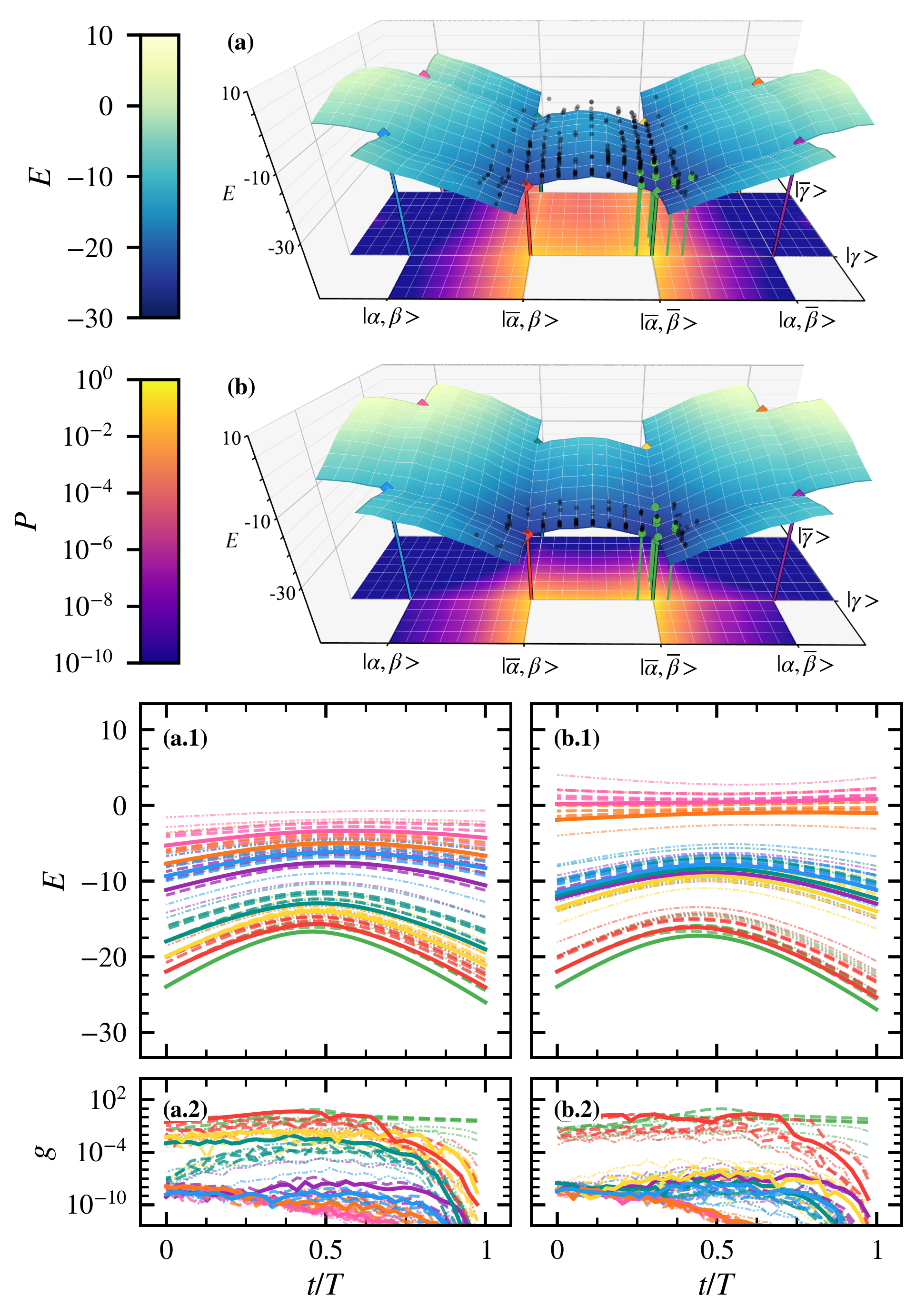}
  \caption{Two iterations of the global search protocol applied to the $24$-qubit spin-glass instance shown in Fig.~\ref{fig:spin_glass_graph}a, using $Q_{\mathcal{S}} = 0.98$, $Q_{\alpha^*} = 0.9$, $N_{\mathcal{T}} = 5$, and $\lambda = 1.0$. (a) and (b) depict the energy landscape ($E$) of $H_{F\mathcal{M}}$ during the first and second iterations, respectively. Black dots represent $2000$ samples from a quantum annealing simulation ($T = 10$), with the probability density ($P$) projected onto the bottom plane. Green dots indicate the subset $\mathcal{T}$. Green, brown and yellow dots indicate $\mathcal{T}$ from the first iteration with $H_{F\mathcal{M}} = H_P$ in Fig.~\ref{fig:spin_glass_graph}c. $\ket{\overline{\alpha}, \beta, \gamma}$ is the initial local minimum $\alpha^*$ (red diamond). The color scales are consistent with those in Fig.~\ref{fig:spin_glass_graph}c.  Panels (a-b.1) display the instantaneous energy spectrum of the transverse field Ising model with $H_{F\mathcal{M}}$ for each iteration. The energy levels correspond to those in Fig.~\ref{fig:energyspectrum_g_orig}, where the color of the lines indicates the association to the eight valleys in the energy landscape of $H_P$. A linear annealing schedule $\Gamma(t)$ is used throughout as an example. Panels (a-b.2) show the adiabatic ratios $g$ (see \equref{eq:g}) for the states shown in panels (a-b.1), respectively.}
  \label{fig:global_search}
\end{figure}

Figure~\ref{fig:global_search} presents two iterations of the global search protocol applied to the $24$-qubit spin-glass instance depicted in Fig.~\ref{fig:spin_glass_graph}a, using parameters $Q_{\mathcal{T}} = 0.98$, $Q_{\alpha^*} = 0.9$, $N_{\mathcal{T}} = 5$ and $\lambda = 1.0$. The state $\left(\overline{\alpha}, \beta, \gamma\right)$ (red diamond) is selected as the initial local minimum $\alpha^*$. In the first iteration (see Fig.~\ref{fig:global_search}a), the protocol identifies $\overline{\alpha}$ as the optimal ground state of the first domain, followed by identifying $\gamma$ as the optimal ground state of the third domain in the second iteration (see Fig.~\ref{fig:global_search}b). As a result, the protocol progressively isolates the valleys corresponding to the states $\left(\overline{\alpha}, \beta, \gamma\right)$ (red curves) and $\left(\overline{\alpha}, \overline{\beta}, \gamma\right)$ (green curves) in the instantaneous energy spectrum (see Fig.~\ref{fig:global_search}b.1), yielding states from the latter (green dots in panel b) with high probability.

\begin{algorithm}
    \caption{Global search}
    \label{algo:global_search}
    \begin{algorithmic}[1]
        \vskip 6pt
        
        \Require {
            \begin{minipage}[t]{0.85\linewidth}
                $\alpha^*$ \Comment{Initial local minimum}\\
                $H_P$ \Comment{Problem Hamiltonian}\\
                $N_{\mathcal{S}}$ \Comment{Size of $\mathcal{S}$}\\
                $N_{\mathcal{T}}$ \Comment{Size of $\mathcal{T}$}\\
                $Q_{\mathcal{T}}$ \Comment{Max. similarity between states in $\mathcal{T}$}\\
                $Q_{\alpha^*}$ \Comment{Max. similarity to $\alpha^*$}\\
                $\lambda$ \Comment{Mixing strength}\\
                $I$ \Comment{Number of iterations}\\
                $T$ \Comment{Annealing time}
            \end{minipage}
        }\vskip 12pt

        \Ensure {
             \begin{minipage}[t]{0.82\linewidth}
                $\beta$ \Comment{State from a lower-energy valley}
            \end{minipage}
        }\vskip 12pt

        \Procedure{Global search}{}\vskip 2pt

            \State $\mathcal{M} \gets 0 \dots 0 0$\vskip 2pt

            \For{$i$ \texttt{in range} (0, I)}\vskip 2pt
            
                \State $H_{F\mathcal{M}} \gets \texttt{generate\_}H_{F\mathcal{M}}\left(H_P, \alpha^*, \mathcal{M}, \lambda\right)$\Comment{Eq.~\ref{eq:H_F_M}}\vskip 2pt
                
                \State $\mathcal{S} \gets \texttt{sample}\left(H_{F\mathcal{M}}, N_{\mathcal{S}}, T\right)$\vskip 2pt
                
                \State $\mathcal{T} \gets \texttt{select\_samples}\left(\mathcal{S}, H_P, N_{\mathcal{T}}, Q_{\mathcal{T}}\right)$\Comment{Eq.~\ref{eq:Tau}}\vskip 2pt

                \State $\mathcal{T} \gets \mathcal{T} + \left[\alpha^*\right]$

                \State $\mathcal{M} \gets \texttt{update\_}\mathcal{M}\left(\mathcal{T}\right)$ \Comment{Eq.~\ref{eq:M}}\vskip 2pt

                \State $\beta \gets \mathcal{T}[0]$\vskip 2pt

                \If{$E_{\beta} < E_{\alpha^*}$}\vskip 2pt
                    \State \Return $\beta_i$\vskip 2pt
                \EndIf
                
            \EndFor

            \State \Return $\beta$\vskip 6pt
        
        \EndProcedure

    \end{algorithmic}
\end{algorithm}


\section{Benchmarks}
\label{sec:benchmarks}
We assess the performance of the proposed hybrid solver with LDA against other quantum and classical methods on large spin-glass instances. We focus on random spin glasses that can be natively mapped onto currently available D-Wave Advantage1 QPUs. Specifically, we investigate $10$ NAT-7~\cite{NAT_7_1, NAT_7_2} instances that almost fully utilize the Advantage1 5.4 QPU~\cite{QPU_Advantage_5_4}, using $5,580$ qubits and $39,898$ couplers. The instances are generated by assigning random coupling strengths $J$ to each coupler, where $J$ is uniformly chosen from the set $\mathcal{J} = \left\{\pm 1/7, \, \pm 2/7, \, \dots \, , \, \pm 1\right\}$. No biases $h$ are applied to the qubits, maintaining a global $\mathbb{Z}_2$ symmetry. Using multiples of $1/7$ ensures that the problem and feature Hamiltonians are well within the precision limits of the QPU~\cite{QPU_ICE}. Moreover, the use of a non-Sidon~\cite{Sidon} set (i.e., the pairwise sum of two elements from the set can be an element of the set again) allows for vanishing fields and causes degeneracies between states. The problem instances are designed to feature numerous low-energy local minima, that have energies within a percent of the ground state energy but are separated by large energy barriers (i.e., $d_{ham} = \mathcal{O}(10^2)$; $\mathcal{O}$ here denotes order of magnitude in the physics convention rather than asymptotic scaling in the computer-science convention). This enables us to investigate the capability of quantum and classical solvers to navigate rough energy landscapes.

\begin{figure*}
  \centering
  \includegraphics[width=\linewidth]{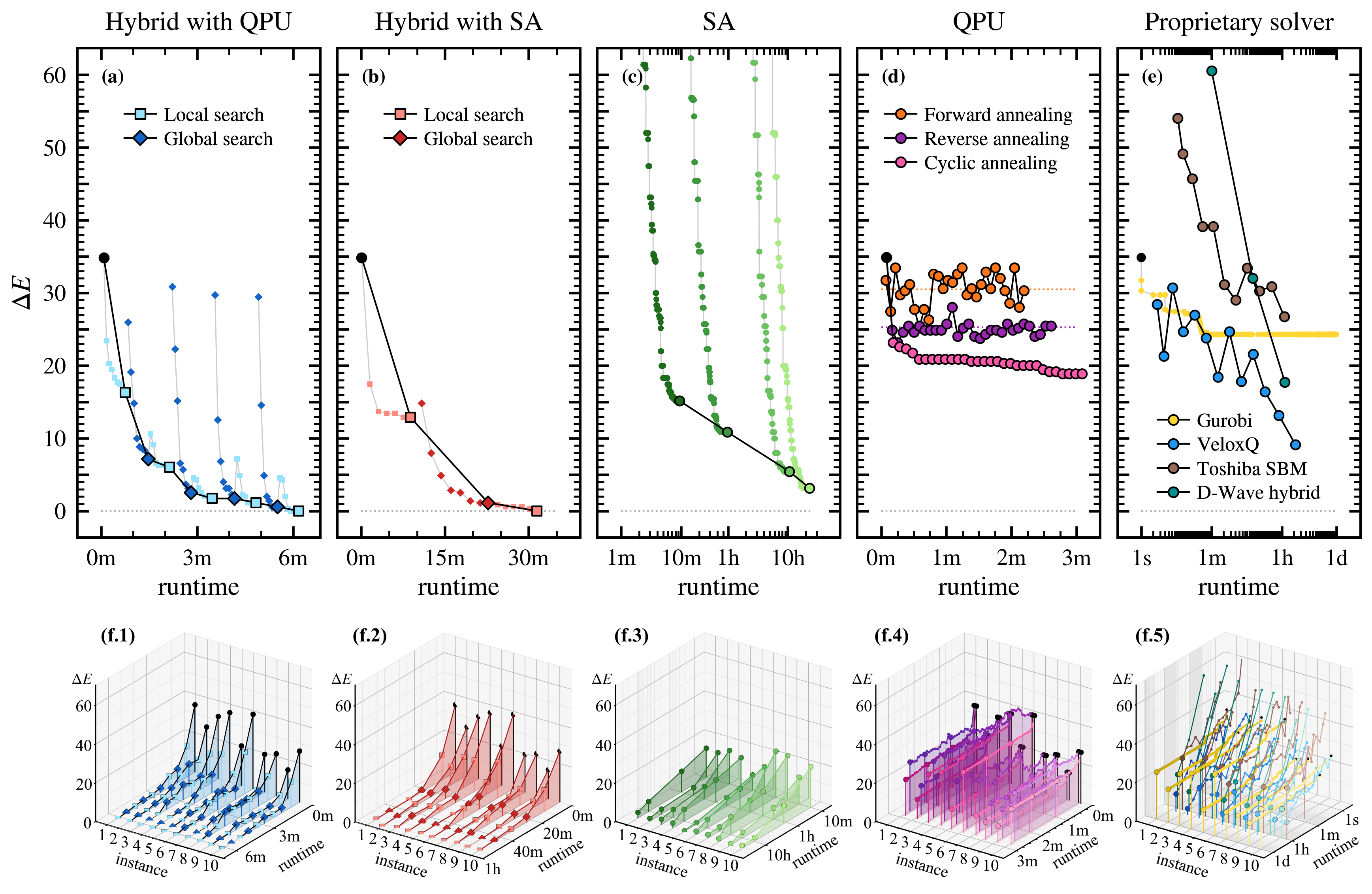}
  \caption{Performance comparison of various quantum and classical solver on $10$ random $5580$-qubit NAT-7 spin-glass instances. (a)--(e) Relative energy to the best solution as a function of algorithm runtime for the first problem instance. Note that only the active QPU and CPU runtimes are reported, thus excluding queue waiting times. Classical code was executed on the JUWELS Booster supercomputer~\cite{Juwels_Cluster}. (a) The proposed hybrid solver using the D-Wave Advantage1 5.4 QPU for sampling ($2ms$ annealing time, spin-reversal transformation, $2000$ samples per run). (b) The proposed hybrid solver using SA from the D-Wave Neal SDK~\cite{D_Wave_NEAL} for sampling ($500,000$ sweeps, $200$ samples, $48$ threads on an Intel Xeon Platinum 8168 CPU). In both panels, squares denote the local protocol ($8$ iterations) and diamonds show the global protocol ($8$ iterations). (c) $4$ executions of SA using JUPTSA~\cite{JUPTSA} (Intel Xeon Platinum 8168 CPU with $48$ threads) and a geometric annealing schedule form $\beta = 0.1$ to $\beta = 10$ with $600K$ ($\approx 10$m), $4M$ ($\approx 1$h), $40M$ ($\approx 10$h), and $96M$ ($\approx 24$h) spin evaluations. (d) Quantum algorithms on the D-Wave Advantage1 5.4 QPU: forward annealing (orange dots, annealing schedule: $\left[(0, 0), (2000, 1)\right]$, $2000$ samples, spin-reversal transformation, mean: orange dotted line); reverse annealing~\cite{Ra_Hg} (purple dots, annealing schedule: $\left[(0, 1), (200, 0.7), (1800, 0.7), (2000, 1)\right]$, \textit{reinitialize\_state=false}, $2000$ samples, spin-reversal transformation, mean: purple dotted line); cyclic annealing~\cite{Cyclic_annealing_1, Cyclic_annealing_2} (pink dots, h-gain schedule: $\left[(0, 0), (1, 0.03), (4, 0.03), (2000, 0)\right]$, annealing schedule: $\left[(0, 1), (1, 1), (4, 0.7), (2000, 1)\right]$, $2000$ samples, spin-reversal transformation). (e) Proprietary/classical solvers: Gurobi~\cite{Gurobi} (yellow dots), D-Wave hybrid~\cite{D_Wave_Hybrid} (teal dots, classified as a classical solver here as the QPU access time was $\lesssim 1\%$ of the total runtime) using $1m$, $10m$ and $1h$ time limits, Toshiba's SBM algorithm~\cite{Toshiba_SBM} implemented in Ref.~\cite{VeloxQ} (brown dots; 4 Nvidia H100 GPUs) and VeloxQ~\cite{VeloxQ, VeloxQ2} (light blue dots; 4 Nvidia H100 GPUs; instance specific tuning of parameters). (f.1-5) Cumulative results of all $10$ NAT-7 instances and solvers. Note, that for each problem instance $i$ the proposed hybrid solver (both QPU and SA sampling), reverse- and cyclic-annealing, and Gurobi use the same initial state $\Psi_i$ (black dot), which is the lowest-energy state obtained from a single forward annealing run on the D-Wave Advantage1 5.4 QPU ($2ms$ annealing time, spin-reversal transformation, $2000$ samples). See App.~\ref{app:qpu_settings} for a detailed description of the QPU settings.}
  \label{fig:benchmark_nat7}
\end{figure*}

Figure~\ref{fig:benchmark_nat7} presents the benchmarking results of various quantum and classical algorithms for the NAT-7 instances, with the relative energy to the best state detailed as a function of runtime for the first problem instance in panels (a)--(e). Results for the remaining instances are summarized in panels (f.1-5) (for reference, comparisons in terms of absolute energies are given in \figref{fig:benchmark_nat7_all} in App.~\ref{app:benchmark}). For each problem instance $i$, a reference state $\Psi_i$ (black dots) is determined, which is the lowest-energy state obtained from a forward annealing run on the D-Wave Advantage1 5.4 QPU, using an annealing time of $2ms$ and generating $2000$ samples. This reference state $\Psi_i$ is used to initialize the proposed hybrid solver and other quantum strategies, to ensure a fair comparison with regard to the use of the QPU to explore the energy landscape. 

Panel~\ref{fig:benchmark_nat7}a shows results of the proposed hybrid solver executed on the D-Wave Advantage1 5.4 QPU. The two stages of the algorithm are distinguished by marker shape: squares denote $8$ iterations of the local search protocol, while diamonds represent $8$ iterations of the global search protocol. Starting from $\Psi_1$, the solver produces $2000$ samples per iteration with an annealing time of $2ms$ and spin-reversal transformations enabled (see App.~\ref{app:qpu_settings}). Examples of sampled states for the two protocols are shown in App.~\ref{app:demonstration_local_global}. The solver achieves the best energy of $-9828.14$ ($\Delta E = 0$) within $4$ cycles, corresponding to a wall-clock runtime of $\approx 6$ minutes. This runtime includes both active CPU time and QPU access time reported by D-Wave, but excludes QPU queue waiting time. The classical processing overhead of LDA amounts to $\approx 49s$, or $\approx 13.5\%$ of the total runtime, due to serial post-processing of the sampled states. Notably, this overhead could be reduced substantially through parallelization.

For comparison, panel~\ref{fig:benchmark_nat7}d presents the lowest energies (orange dots) from $30$ independent runs on the QPU for the unmodified problem Hamiltonian using the same annealing parameters as before. The data shows that the QPU samples a mean energy of $\overline{\Delta E} \approx 30.51$ (see dotted orange line) with a root-mean-square deviation of $\sigma_{\Delta E} \approx 2.18$. Already the first iteration of the local search (first square in (a)) is significantly below this energy, with $\Delta E \approx 23.43$. This demonstrates that LDA effectively guides the system through the instantaneous energy spectrum by focussing the dynamics on critical transitions, i.e., avoided crossings with small gaps, thereby allowing the system to repeatedly attempt traversing them. As a result, LDA gradually eliminates avoided crossings from the spectrum and thereby reach deeper regions of the energy landscape that are inaccessible within the annealing time of the QPU.

While LDA is primarily designed to suppress QPTs during QA, the gradual simplification of the energy landscape can also benefit classical sampling techniques such as simulated annealing (SA), TABU search, and evolutionary algorithms. To explore this potential, we replace QPU sampling in our hybrid solver with SA in Panel~\ref{fig:benchmark_nat7}b. For convenience, we employ the SA implementation from the D-Wave Neal SDK~\cite{D_Wave_NEAL}, generating $200$ samples with $500,000$ sweeps on $48$ threads of an Intel Xeon Platinum 8168 CPU, initialized from $\Psi_1$ (black dot). With SA, the hybrid solver identifies the best energy within a single cycle in $\approx 30min$. Although classical sampling with SA is considerably slower than QPU sampling (noting that runtime could be reduced with more computational ressources), this result demonstrates that LDA is applicable beyond QA. Moreover, other classical sampling methods combined with LDA may converge even faster to the solution state, though we leave a detailed study on classical sampling for future work.

In the context of QA, the success of the SA-based hybrid solver highlights a general property of LDA: the ability to reach the GS does not depend on the quality of the sampled states. Lower-quality samples merely increase the number of cycles required to reach the GS. In this experiment, the SA-based version converged in fewer cycles than the QPU version, as it produced lower-energy samples. This indicates that LDA is broadly applicable to a wide range of NISQ devices, where QA is limited by integrated control errors and finite annealing times~\cite{QPU_Advantage_5_4, QPU_ICE}, since convergence speed (i.e., number of cycles) naturally adapts to the available coherence time.

In panel~\ref{fig:benchmark_nat7}c, we present results from conventional SA using a parallel SA software called JUPTSA~\cite{JUPTSA} with $48$ threads. A total of $4$ runs was conducted with a geometric annealing schedule from $\beta = 0.1$ to $\beta = 10$, and $600K$ ($\approx 10$m), $4M$ ($\approx 1$h), $40M$ ($\approx 10$h), and $96M$ ($\approx 24$h) spin evaluations. Notably, with an energy of $\Delta E \approx 3.14$, SA does not reach the best energy within the $24h$ time limit, emphasizing the difficulty of the problem instance. Compared to panels (a) and (b), our hybrid solver finds the final energy of the $24h$ run in $168.13s$ and $961.45s$ using QPU and SA sampling, respectively. This speed-up is the result of the proposed local and global search protocols, and highlights their efficiency in navigating the rough energy landscape and locating low-energy valleys. The comparison of panels (b) and (c) also reveals that short annealing runs are better for LDA, as it gives the algorithm greater control over the time-evolution to focus the dynamics into promising regions of the Hilbert space.

This observation is further supported by the results of alternative quantum algorithms shown in panel~\ref{fig:benchmark_nat7}d, which make use of advanced programming features of D-Wave QPUs. The first ansatz we examine is reverse annealing using the IRA scheme~\cite{Ra_Hg} as discussed in Sec.~\ref{sec:qa} (purple dots). Initializing the system in the state $\Psi_1$ (black dot), we execute in total $30$ runs on the QPU, with \textit{reinitialize\_state=false} and $\left[(0, 1), (200, 0.7), (1800, 0.7), (2000, 1)\right]$ as the annealing schedule (see App.~\ref{app:qpu_settings}). Each dot denotes the lowest-energy state of $2000$ samples from each iteration. The data shows that reverse annealing improves on the initial energy, but fails to reach the best energy with a mean of $\overline{\Delta E} \approx 25.26$ (see dotted purple line) with a root-mean-square deviation of $\sigma_{\Delta E} \approx 1.94$. As discussed in Sec.~\ref{sec:qa}, this is likely because reverse annealing is constrained by the QPU's effective temperature. 

The second strategy we investigate termed cyclic annealing (pink dots) was proposed by Wang et al.~\cite{Cyclic_annealing_1, Cyclic_annealing_2} and combines reverse annealing with the \texttt{h\_gain} schedule feature (which controls the function $b(t)$ in \equref{eq:QPUHamiltonian}, see App.~\ref{app:qpu_settings}) to cycle around the tricritical point of the many-body localization transition~\cite{SG_QPT_Theory}. This is achieved using a reference Hamiltonian that encodes the initial state of each cycle through local biases.~The strength of this reference Hamiltonian is controlled via the \texttt{h\_gain\_schedule} $\left[(0, 0), (1, 0.05), (40, 0.05), (2000, 0)\right]$.~The reverse annealing schedule is $\left[(0, 1), (1, 1), (40, 0.4), (2000, 1)\right]$.~Although cyclic annealing achieves lower energies than reverse annealing alone (best energy at $\Delta E \approx 20.90$), we observe that each cycle results in only minor changes to the bit string, with $\mathcal{O}(10^1)$ Hamming distance between initial and final state (for comparison, the proposed global search protocol achieves $\mathcal{O}(10^2)$ bit flips). As a result, cyclic annealing often becomes trapped in local minima. This behavior likely arises from the design of the reference Hamiltonian, which biases states based on their Hamming distance to the initial state, restricting the search to nearby states. Given that energy barriers can span $\mathcal{O}(10^2)$ bits, cyclic annealing is unlikely to transition between energy valleys, instead requiring exponentially many restarts to locate the GS.

As a result, all tested purely QPU-based strategies fail to find the best energy on the D-Wave Advantage1 5.4 QPU. While future generations of QPUs might be able to natively sample the best energy, our SA results (panels b, c) indicate that a hybrid protocol, modifying the problem Hamiltonian according to \equref{eq:H_F_M}, might still be the best choice for practical quantum computation, as it can guide the evolution more effectively (panel a).

As our final benchmark, we consider proprietary/classical solvers in panel~\ref{fig:benchmark_nat7}e. These include Gurobi~\cite{Gurobi} (yellow dots), D-Wave hybrid~\cite{D_Wave_Hybrid} (teal dots), and two quantum-inspired-algorithms (QIA): Toshiba's simulated bifurcation machine (SBM)~\cite{Toshiba_SBM} implemented in Ref.~\cite{VeloxQ} (brown dots), and VeloxQ~\cite{VeloxQ, VeloxQ2} (light blue dots). Note that QIA are often regarded as a classical baseline for quantum algorithms~\cite{QAIA}. For Gurobi, we conducted two executions: one without an initial state and one initialized with the state $\Psi_1$ (black dot). Both executions had a $24h$ time limit. Only the results of the second run are shown, as the first run failed to achieve an energy lower than $\Delta E \approx 474.57$. Using the initial state, Gurobi improves within the first minute, reaching an energy of $\Delta E \approx 24.2857$. However, subsequent progress was minimal, with the only improvements occurring to the lower-bound energy of the ground state, such that the best energy was not reached within the $24h$ runtime. Regarding D-Wave hybrid, we evaluated runtimes of $1min$, $10min$, and $1h$, with the latter reaching an energy of $\Delta E \approx 17.71$. Throughout these runs, QPU access time was $\lesssim 1\%$ of the total runtime, suggesting the solver operated predominantly as a classical solver. The poor QPU utilization is likely due to the difficulty of finding a suitable embedding for the problem. Toshiba's SBM and VeloxQ achieve energies of $\Delta E \approx 26.72$ and $\Delta E \approx 9.14$, respectively. Note that both QIA solvers were executed on substantially more powerful classical hardware (four Nvidia H100 GPUs) compared to the other classical solvers (Intel Xeon Platinum 8168 CPU). We remark that the lowest energies were found by SA and our hybrid solver, while all investigated proprietary classical solvers would likely require longer runtimes or more computing resources to reach similar energies.

In panel (f), we present the cumulative results for the $10$ problem instances. Across all instances, we observe consistent qualitative behavior for the evaluated algorithms. Notably, the proposed hybrid solver (both the QPU and SA implementation) finds the lowest energy across all investigated problem instances, surpassing the other algorithms in both lowest energy and total runtime. Only for the fifth and ninth spin-glass instance, SA is able to match the energy of the proposed hybrid solver in the $24$h run.


\section{Conclusion}
\label{sec:conclusion}
In this work, we have introduced Learning-Driven Annealing (LDA), a framework designed to link individual quantum annealing (QA) runs into a global solution strategy to mitigate hardware constraints such as finite annealing times and integrated control errors. LDA differs from other iterative strategies in that it does not change the annealing procedure (i.e, annealing schedule or annealing time). Instead, it learns about the problem structure (i.e.~the spin domains) to adaptively modify the problem Hamiltonian $H_P$. LDA replaces $H_P$ with a so-called feature Hamiltonian $H_{F}$ to iteratively eliminate QPTs in the instantaneous energy spectrum. $H_{F}$ is defined w.r.t.~a reference state $\alpha$, retaining only the biases and couplers from $H_P$ that are satisfied by $\alpha$. These retained terms, called features, encode the location of a nearby local minimum $\alpha^*$ in the energy spectrum. 

We have demonstrated that $H_F$ deforms the energy landscape according to both the Hamming distance and the energy distance from $\alpha$. When combined with a bitmask $\mathcal{M}$ that restricts the application of the feature Hamiltonian to a subset of spins, the time evolution through the instantaneous energy spectrum is partially set, allowing LDA to concentrate the dynamics on critical transitions with small energy gaps. This is achieved by gradually removing competing valleys from the energy landscape. In this way, LDA iteratively eliminates avoided crossings to access states deep within the energy valley that would otherwise remain inaccessible within the QPU’s annealing time.  

To demonstrate the efficacy of LDA, we have developed a hybrid quantum-classical solver for large-scale spin-glass problems. The solver alternates between a local and global search protocol, both using an iterative application of LDA.

The local search protocol converges from an initial state $\alpha_0$ to a nearby local minimum $\alpha^*$. It begins with a distribution phase that scatters the system around the initial state to identify characteristic features of the energy valley, followed by a convergence phase that gradually narrows the search area to isolate $\alpha^*$. Numerical simulations on a $24$-qubit spin-glass confirmed that this approach isolates the energy valley of $\alpha^*$ in the instantaneous energy spectrum, ensuring that the search is restricted to transitions within the valley.

The global search protocol transitions from a local minimum $\alpha^*$ to a state in a lower-energy valley by leveraging the hierarchy of local minima. Using $\alpha^*$ as a reference, the protocol identifies spin domains in suboptimal configurations and gradually replaces the problem Hamiltonian $H_P$ with the feature Hamiltonian $H_F$. Through numerical simulation, we verified that this process isolates lower-energy valleys in the instantaneous energy spectrum, allowing the system to cross high and wide energy barriers and reach a lower-energy valley even at large Hamming distances.

We have benchmarked the proposed hybrid solver against leading quantum and classical methods, including reverse annealing~\cite{IRA_Theory_1}, cyclic annealing~\cite{Cyclic_annealing_1, Cyclic_annealing_2}, simulated annealing~\cite{JUPTSA}, Gurobi~\cite{Gurobi}, Toshiba's SBM~\cite{Toshiba_SBM}, D-Wave hybrid~\cite{D_Wave_Hybrid} and VeloxQ~\cite{VeloxQ, VeloxQ2}, on $5580$-qubit NAT-7~\cite{NAT_7_1, NAT_7_2} spin-glass instances. The results show that the proposed hybrid solver outperforms all competing algorithms in both runtime and lowest energy using both QPU sampling from the D-Wave Advantage1 5.4 and SA sampling. Remarkably, none of the other methods matched the best energies found by the proposed hybrid solver within a $24h$ time limit (except for SA in the fifth and ninth problem instance), effectiveness of our protocols in exploring rugged energy landscapes and identifying low-energy valleys, even for large-scale instances. 

It is important to emphasize that these results do not demonstrate a quantum advantage, as the same energies obtained with QPU sampling can also be reproduced by combining LDA with classical SA. Nevertheless, our experiments on real quantum hardware show that LDA effectively mitigates hardware limitations of NISQ devices (particularly short coherence times) by systematically guiding the time evolution through the instantaneous spectrum. This indicates that, in principle, NISQ devices could solve any COP that can be embedded onto the hardware using LDA, although further studies are needed to assess the applicability of our local and global search protocols to other problem classes. Moreover, LDA could be integrated with hybrid quantum–classical sampling strategies (e.g., combining classical local search with QPU-based global search), potentially enabling even faster convergence. Overall, LDA represents a step toward practical quantum optimization by providing classical control over the quantum time evolution, allowing current quantum annealers to compete with state-of-the-art classical solvers in both runtime and solution quality.


\begin{acknowledgments}
The authors thank Hans De Raedt, Madita Willsch, Vrinda Mehta, Fengping Jin, Jaka Vodeb, and Paul Warburton for comments and discussions. The authors thank 
Jakub Pawłowski and Bartłomiej Gardas for sharing performance results of Toshiba's SBM and VeloxQ on the investigated problem instances. The authors gratefully acknowledge the Gauss Centre for Supercomputing e.V. (www.gauss-centre.eu) for funding this project by providing computing time on the GCS Supercomputer JUWELS~\cite{Juwels_Cluster} at Jülich Supercomputing Centre (JSC). The authors gratefully acknowledge the Jülich Supercomputing Centre (https://www.fz-juelich.de/ias/jsc) for funding this project by providing computing time on the D-Wave Advantage™ System JUPSI through the Jülich UNified Infrastructure for Quantum computing (JUNIQ). D.W. acknowledges support from the project JUNIQ that has received funding from the German Federal Ministry of Education and Research (BMBF) and the Ministry of Culture and Science of the State of North Rhine-Westphalia.
\end{acknowledgments}


\bibliography{bibliography}
\clearpage
\appendix

\section{Asymmetry of $q_F$}
\label{app:asymmetry_qf}
\begin{figure*}
  \centering
  \includegraphics[width=0.8\linewidth]{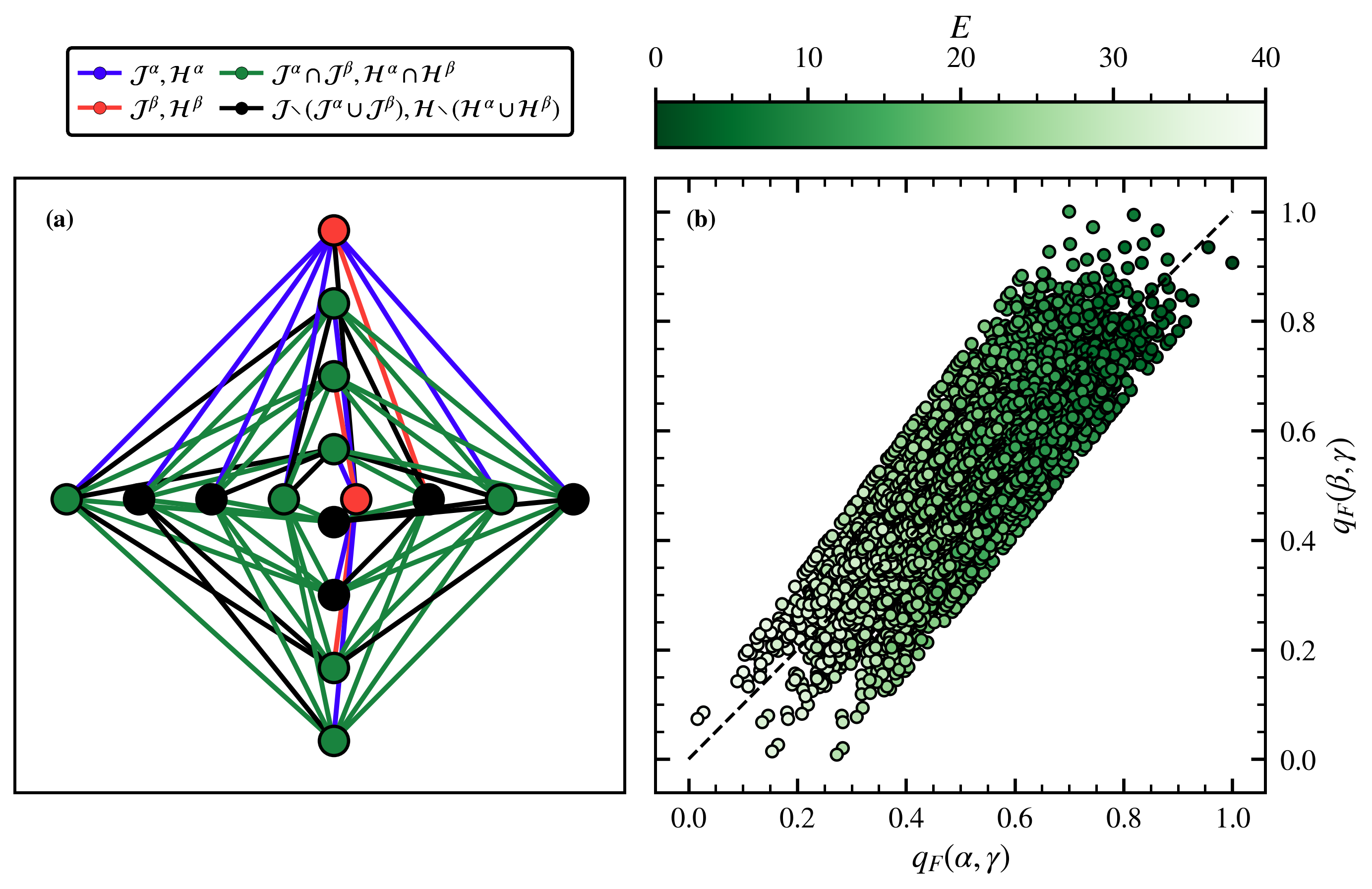}
  \caption{Asymmetry of the similarity measure $q_F$ w.r.t.~the global minimum $\alpha$ and a state $\beta$ located in the global minimum well. (a) $16$-qubit spin-glass graph, where nodes represent biase $h_i$ and edges represent couplers $J_{ij}$. Colors indicate whether $h_i$ and $J_{ij}$ belong to the sets of couplers $\mathcal{J}^{\alpha}, \mathcal{J}^{\beta}$ and biases $\mathcal{H}^{\alpha}, \mathcal{H}^{\beta}$ satisfied by $\alpha$ and/or $\beta$. (b) Comparison of $q_F(\alpha, \gamma)$ and $q_F(\beta, \gamma)$ across all computational basis states $\gamma$, illustrating the asymmetry of the similarity measure. While $q_F(\beta, \alpha)$ is close to one, while $q_F(\alpha, \beta)$ is significantly lower, revealing an intrinsic bias toward the lower-energy state.}
  \label{fig:qf_asym}
\end{figure*}

Figure~\ref{fig:qf_asym} compares the similarity measure $q_F$ for two reference states $\alpha$ and $\beta$ on a $16$-qubit spin-glass instance with four energy valleys. Here, $\alpha$ corresponds to the global minimum, while $\beta$ is a state located in the same energy valley. Both states have a Hamming distance of $d_{ham} = 2$. The problem graph shown in panel (a) reveals that the satisfied terms $\mathcal{J}^{\alpha}, \mathcal{J}^{\beta}$ and $\mathcal{H}^{\alpha}, \mathcal{H}^{\beta}$ coincide for most couplers and biases, indicated in green. State $\alpha$ satisfies five couplers not satisfied by $\beta$ (shown in blue), whereas $\beta$ satisfies three additional couplers and two biases (shown in red). Panel (b) depicts $q_F$ for both reference states across all computational basis states $\gamma$, highlighting the asymmetry of the similarity measure. Notably, while the global minimum (at $q_F(\alpha, \gamma) = 1$) also yields $q_F(\beta, \gamma) \approx 1$, the reverse overlap $q_F(\alpha, \beta)$ is significantly smaller. This demonstrates that $q_F$ is intrinsically biased toward lower-energy states. As a result, the feature Hamiltonian on average guides the system down an energy valley.

\section{QPU settings}
\label{app:qpu_settings}
The time-dependent Hamiltonian implemented on the D-Wave Advantage1 5.4 QPU is defined as
    \begin{align}
        H_{QA}(t) = &- \frac{A\left(s(t)\right)}{2} \left[\sum_i \sigma_x^i\right] \\
        \label{eq:QPUHamiltonian}
        &+ \frac{B\left(s(t)\right)}{2} \left[b(t) \cdot \sum_i h_i \sigma_z^i + \sum_{i < j} J_{ij} \sigma_z^i \sigma_z^j\right],
    \end{align}
where $h_i$ and $J_{ij}$ represent the linear and quadratic terms of the problem Hamiltonian $H_P$, respectively. The function $s(t) \in \left[0, 1\right]$ parameterizes the annealing schedule for which $A$ and $B$ are the strengths of the driving and problem Hamiltonians, respectively (the particular functions $A(s)$ and $B(s)$ for the QPU used are shown in \cite{QPU_Advantage_5_4}). The function $b(t)$ is called \texttt{h\_gain} schedule. 

For forward annealing, we use $s(t) = t/T$ and $b(t) = 1$, where $T$ is the total annealing time. Additionally, we apply a spin-reversal transformation to $H_P$ for every batch of $400$ samples from the QPU. This transformation involves generating a random binary string $r \in \left\{\pm 1\right\}^N$ and modifying the problem terms via $h_i \xrightarrow{} h_i r_i$ and $J_{ij} \xrightarrow{} J_{ij} r_i r_j$. Sampled states $\alpha$ are transformed back using $\alpha_i \xrightarrow{} \alpha_i r_i$. We use the spin-reversal transformation to mitigate memory effects on the QPU when rapidly submitting similar problems.

Reverse annealing is realized by changing the annealing schedule via $s(t)$ to begin at $s(0) = 1$ where $B(s) \gg A(s)$. The schedule is specified as a series of tuples $[t/T, s]$ (see~\secref{sec:benchmarks}), with linear interpolation between points. Inspired by \cite{IRA_Failing_1}, we use a schedule with a symmetric pause, by first annealing to an intermediate value $0 < s'< 1$, holding $s'$, and then returning to $s=1$. For the large spin glasses under investigation, we found $s' = 0.7$ to be optimal. For all reverse annealing runs, we set \textit{reinitialize\_state=false}, initializing each run from the previously sampled state.

Cyclic annealing uses both reverse annealing and the QPU's \texttt{h\_gain} feature, which introduces a time-dependent gain $b(t)$ for the linear terms $h_i$ in $H_P$. Similar to $s(t)$, $b(t)$ is defined as series of tuples $[t/T, b]$, with linear interpolation between points. Since the large spin glasses under investigation have no linear biases (i.e., $h_i = 0$), we use these terms to encode the initial state $\alpha$ of each cycle, setting $h_i = -\alpha_i$. The schedules for $s(t)$ and $b(t)$ are inspired by \cite{Cyclic_annealing_2}.

\section{Properties of $H_F$}
\label{app:prop_Hf}

\begin{theorem}
Given the feature Hamiltonian $H_F(\alpha)$ (see \equref{eq:H_F}) to a reference state $\alpha$, then $\alpha$ is a ground state of $H_F(\alpha)$.
\label{lem:gs}
\end{theorem}

\begin{proof}
Let $H_F(\alpha)$ be the feature Hamiltonian to a reference state $\alpha = \alpha_{N-1} \dots \alpha_1 \alpha_0$ and let $\beta = \beta_{N-1} \dots \beta_1 \beta_0$ be an arbitrary state, with $\alpha_i, \beta_i \in \left\{-1, 1\right\}$. Then
    \begin{align}
        \label{eq:appA1}
        E_{\beta} &= \bra{\beta}H_F(\alpha)\ket{\beta}\\
        \label{eq:appA2}
        &= \;\;\sum\limits_{(i,j)\in \mathcal{J}^{\alpha}}\frac{-\abs{J_{ij}}}{2} \cdot \left[\alpha_i \alpha_j \beta_i \beta_j + \alpha_i \beta_i + \alpha_j \beta_j\right]\\
        \label{eq:appA3}
        &\;\quad + \sum\limits_{i\in \mathcal{H}^{\alpha}} h_i \beta_i.
    \end{align}
For each term in the first sum, we have
\begin{align}
    \alpha_i \alpha_j \beta_i \beta_j &+ \alpha_i \beta_i + \alpha_j \beta_j \nonumber\\
    \label{eq:appalphabeta}
    &= \begin{cases}
        3 & \text{if }(\beta_i,\beta_j)=(\alpha_i,\alpha_j) \\
        -1 & \text{if }(\beta_i,\beta_j)=(\overline{\alpha_i},\alpha_j) \\
        -1 & \text{if }(\beta_i,\beta_j)=(\alpha_i,\overline{\alpha_j}) \\
        -1 & \text{if }(\beta_i,\beta_j)=(\overline{\alpha_i},\overline{\alpha_j})
    \end{cases},
\end{align}
so \equref{eq:appA2} is minimal if $(\beta_i,\beta_j)=(\alpha_i,\alpha_j)$ for all $(i,j)\in \mathcal{J}^{\alpha}$. Similarly, for the second sum in \equref{eq:appA3}, we have
    \begin{align}
        h_i \beta_i < 0 \Leftrightarrow \beta_i = \alpha_i,
    \end{align}
by the definition of $\mathcal{H}^{\alpha}$ (see~\equref{eq:satisfiedH})
Thus, the energy $E_{\beta}$ is minimal if $\beta = \alpha$. Therefore, $\alpha$ is a ground state of $H_F(\alpha)$.
\end{proof}

\begin{theorem}
Given the feature Hamiltonian $H_F(\alpha)$ (see \equref{eq:H_F}) to a reference state $\alpha$, then the energy $E_\beta$ of an arbitrary state $\beta\neq\alpha$ (see \equref{eq:appA1}) is monotonic as a function of Hamming distance $d_{ham}(\alpha,\beta)$ from $\alpha$. This means that there exists a path from $\beta$ to $\alpha$ via $d_{ham}(\alpha,\beta)$ successive single bit flips, each of which either maintains or decreases the energy.
\label{lem:monotonic}
\end{theorem}

\begin{proof}
Let $H_F(\alpha)$ be the feature Hamiltonian to a reference state $\alpha = \alpha_{N-1} \dots \alpha_1 \alpha_0$, and let $\beta = \beta_{N-1} \dots \beta_1 \beta_0$ be an arbitrary state such that $\alpha \neq \beta$, with $\alpha_i, \beta_i \in \left\{-1, 1\right\}$. To prove the statement, we construct a path $P = (\alpha=\gamma^{(0)}, \gamma^{(1)}, \ldots, \gamma^{(d_{ham}(\alpha,\beta)-1)}=\beta)$ with monotonically increasing energies. 

Let $\gamma^{(0)} = \gamma_{N-1}^{(0)} \dots \gamma_1^{(0)} \gamma_0^{(0)} = \alpha$. Then by Lem.~\ref{lem:gs}, $\gamma^{(0)}$ is a ground state of $H_F(\alpha)$, with all terms in $E_{\gamma^{(0)}}$ giving only negative contributions. To construct the path $P$, we define the step from $\gamma^{(n)}$ to $\gamma^{(n+1)}$ for $n=0,\ldots,d_{ham}(\alpha,\beta)-2$.
The energy of $\gamma^{(n)}$ is, according to \equref{eq:appalphabeta},
    \begin{align}
        E_{\gamma^{(n)}}
        \label{eq:appA6}
        &= \;\;\sum\limits_{\substack{(i,j)\in \mathcal{J}^{\alpha}\\(\gamma^{(n)}_i,\gamma^{(n)}_j)=(\alpha_i,\alpha_j)}}\frac{-3\abs{J_{ij}}}{2}\\
        \label{eq:appA7}
        &+ \;\;\sum\limits_{\substack{(i,j)\in \mathcal{J}^{\alpha}\\(\overline{\gamma^{(n)}_i},\gamma^{(n)}_j)=(\alpha_i,\alpha_j)}}\frac{\abs{J_{ij}}}{2}\\
        \label{eq:appA8}
        &+ \;\;\sum\limits_{\substack{(i,j)\in \mathcal{J}^{\alpha}\\(\gamma^{(n)}_i,\overline{\gamma^{(n)}_j})=(\alpha_i,\alpha_j)}}\frac{\abs{J_{ij}}}{2}\\
        \label{eq:appA9}
        &+ \;\;\sum\limits_{\substack{(i,j)\in \mathcal{J}^{\alpha}\\(\overline{\gamma^{(n)}_i},\overline{\gamma^{(n)}_j})=(\alpha_i,\alpha_j)}}\frac{\abs{J_{ij}}}{2}\\
        \label{eq:appA10}
        &\;\quad + \sum\limits_{i\in \mathcal{H}^{\alpha}} h_i \gamma^{(n)}_i.
    \end{align}
Let $i\in\{0,\ldots,N-1\}$ be the first bit for which $\gamma_i^{(n)} \neq \beta_i$ and set $\gamma^{(n+1)}$ to be the state with the single bit $i$ flipped, i.e., $\gamma^{(n+1)}=\gamma_{N-1}^{(n)}\dots\gamma_{i+1}^{(n)}\beta_i\gamma_{i-1}^{(n)}\dots\gamma_{0}^{(n)}$. If bit $i$ occurs in the set $\mathcal{J}^{\alpha}$ (see \equref{eq:satisfiedJ}), this single bit flip will either increase the energy of the coupling contribution by moving terms from \equref{eq:appA6} to \equref{eq:appA7} or \equref{eq:appA8}---thereby increasing the energy by multiples of $2\abs{J_{ij}}$---or it will maintain the energy by moving terms from \equref{eq:appA7} or \equref{eq:appA8} to \equref{eq:appA9}. Similarly, if bit $i$ occurs in the set $\mathcal{H}^{\alpha}$ (see \equref{eq:satisfiedH}), the bit flip will increase the energy of the bias contribution in \equref{eq:appA10} by $2\abs{h_i}$. Thus each step from $\gamma^{(n)}$ to $\gamma^{(n+1)}$ for $n=0,\ldots,d_{ham}(\alpha,\beta)-2$ will either increase or maintain the energy. After $N$ iterations, $\gamma^{(d_{ham}(\alpha,\beta)-1)} = \beta$ by the definition of the Hamming distance. Therefore, $P$ is a path of states with monotonically increasing energies.
\end{proof}

\section{Illustration of the local and global search protocol}
\label{app:demonstration_local_global}

\begin{figure*}
\centering
\begin{minipage}[t]{.45\textwidth}
  \includegraphics[width=0.9\linewidth]{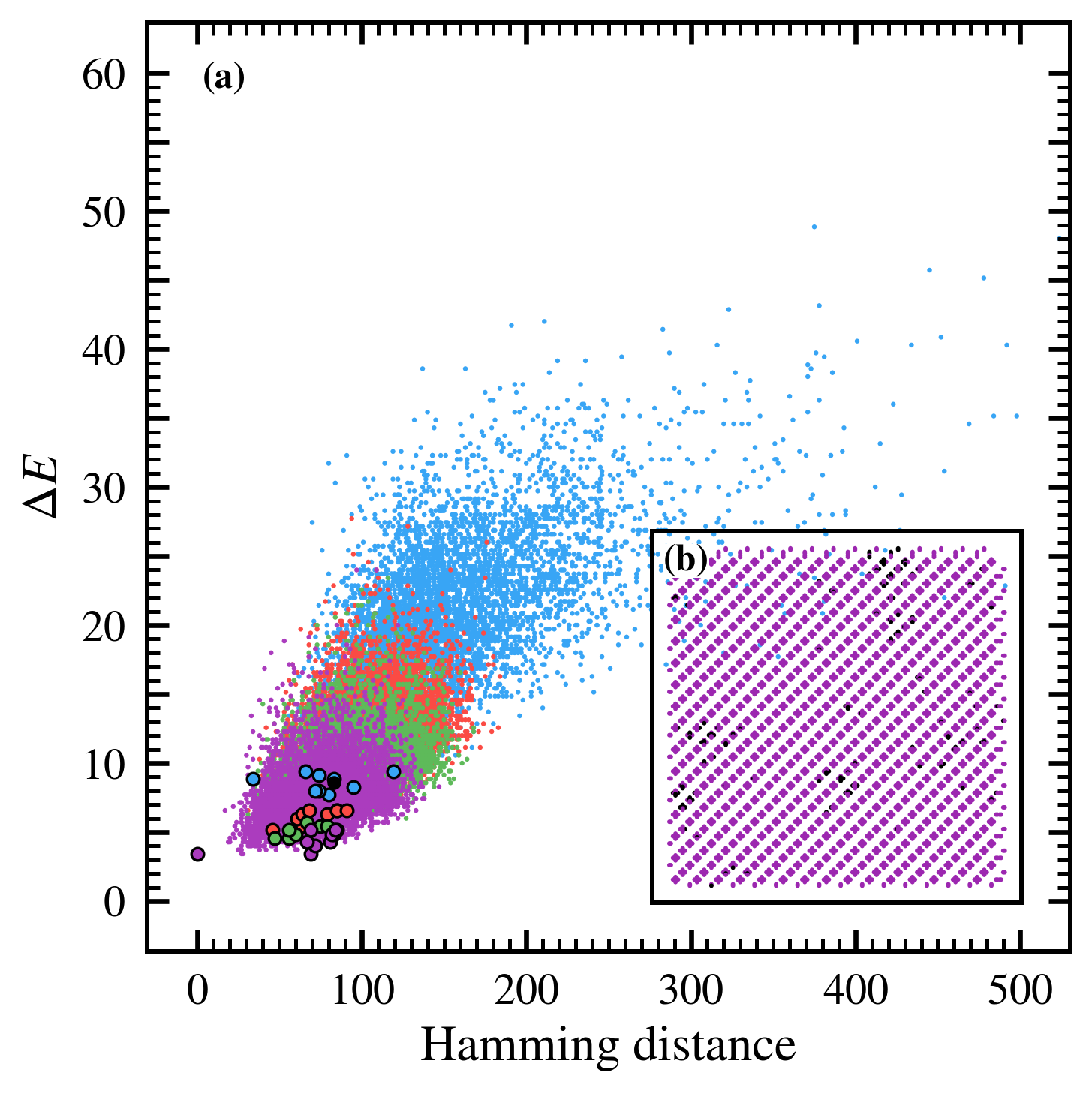}
  \caption{Four iterations of the local search protocol applied to the first NAT-7 spin-glass instance, shown in \figref{fig:benchmark_nat7}. (a) Set $\mathcal{S}$ of $6000$ sampled states for each iteration, presented as a function of their Hamming distance from the protocol's final state (purple circle) and their energy difference $\Delta E$ relative to the best-known state. The energy axis is consistent with \figref{fig:benchmark_nat7}a. The black circle denotes the reference state for the protocol, while blue, red, green, and purple dots represent sampled states from the first, second, third, and fourth iterations, respectively. Colored circles depict the elements of the subset $\mathcal{S}_{\subset}$ used for the calculation of the bitmask $\mathcal{M}$ (see \equref{eq:M}) of each iteration. (b) Embedding of the feature Hamiltonian on the D-Wave Advantage1 5.4 QPU~\cite{QPU_Advantage_5_4}, with qubits represented as nodes. Purple nodes indicate qubits for which a coupler and/or bias was modified in the feature Hamiltonian in the fourth iteration relative to the original problem Hamiltonian. Black nodes represent qubits for which the original problem Hamiltonian was used.}
  \label{fig:local_search_eh}
\end{minipage}\qquad
\begin{minipage}[t]{.45\textwidth}
  \includegraphics[width=0.9\linewidth]{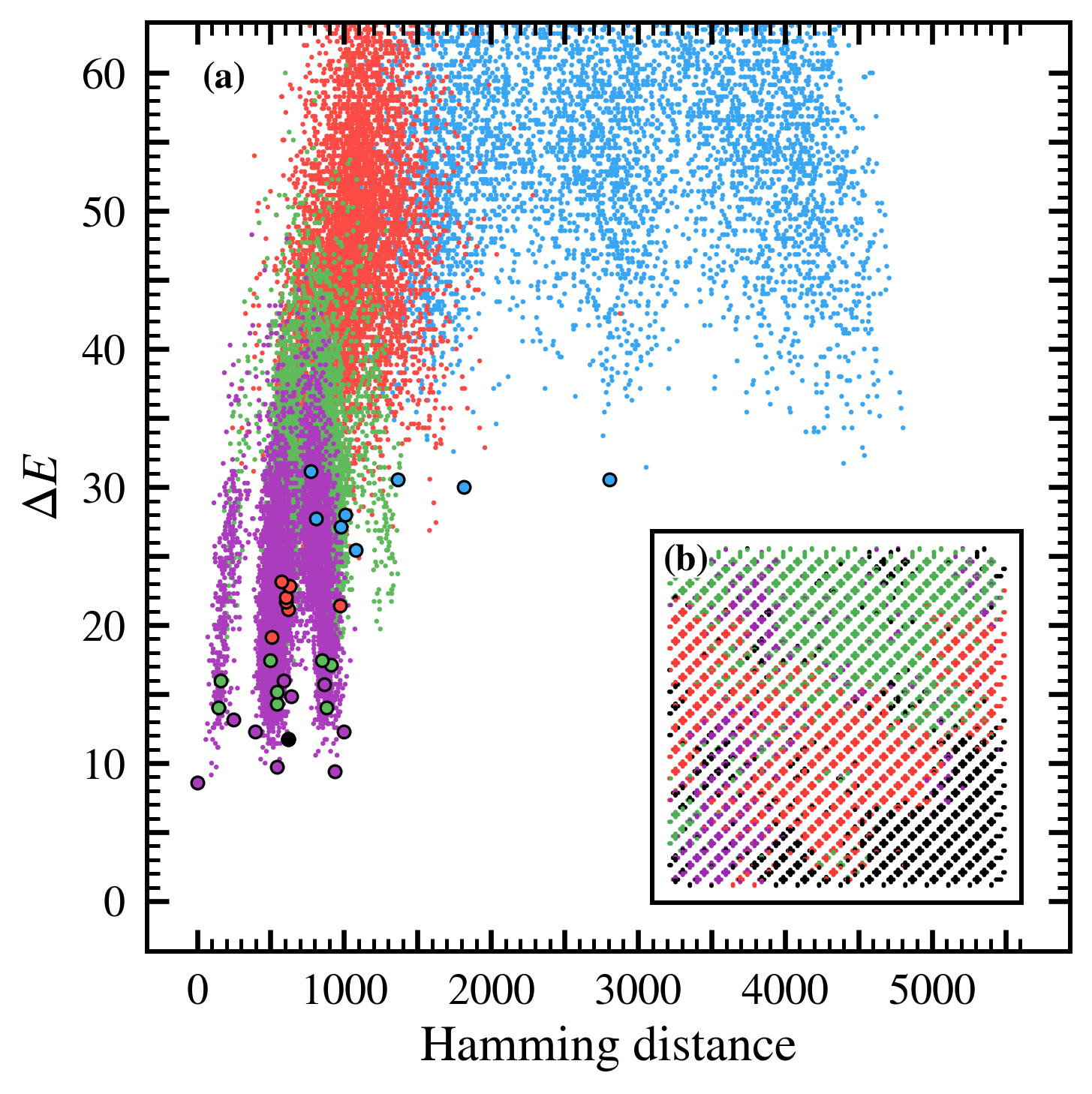}
  \caption{Four iterations of the global search protocol applied to the first NAT-7 spin-glass instance, shown in \figref{fig:benchmark_nat7}. (a) Set $\mathcal{S}$ of $6000$ sampled states for each iteration, presented as a function of their Hamming distance from the protocol's final state (purple circle) and their energy difference $\Delta E$ relative to the best-known state. The energy axis is consistent with \figref{fig:benchmark_nat7}a. The black circle denotes the reference state for the protocol, while blue, red, green, and purple dots represent sampled states from the first, second, third, and fourth iterations, respectively. Colored circles depict the elements of the subset $\mathcal{S}_{\subset}$ used for the calculation of the bitmask $\mathcal{M}$ (see \equref{eq:M}) of each iteration. (b) Embedding of the feature Hamiltonian on the D-Wave Advantage1 5.4 QPU~\cite{QPU_Advantage_5_4}, with qubits represented as nodes. Colored nodes indicate qubits for which a coupler and/or bias was modified in the feature Hamiltonian relative to the original problem Hamiltonian, with the color encoding the iteration in which the modification occurred. Black nodes represent qubits for which the original problem Hamiltonian was used. Since connected qubits are typically spatially close on the QPU, the spin-glass domain structure is reflected in the clustering of colored nodes.}
  \label{fig:global_search_eh}
\end{minipage}

\end{figure*}

Figures~\ref{fig:local_search_eh} and~\ref{fig:global_search_eh} show representative illustrations of the local and global search protocols applied to the first $5580$-qubit NAT-7 spin-glass instance (see \figref{fig:benchmark_nat7}), respectively. They depict the set of sampled states $\mathcal{S}$, the set of selected states $\mathcal{T}$ and the feature Hamiltonian $H_{F \mathcal{M}}$ across four iterations of the protocols.

\section{Benchmarking results of NAT-7 instances}
\label{app:benchmark}
\begin{figure*}
  \centering
  \includegraphics[width=\linewidth]{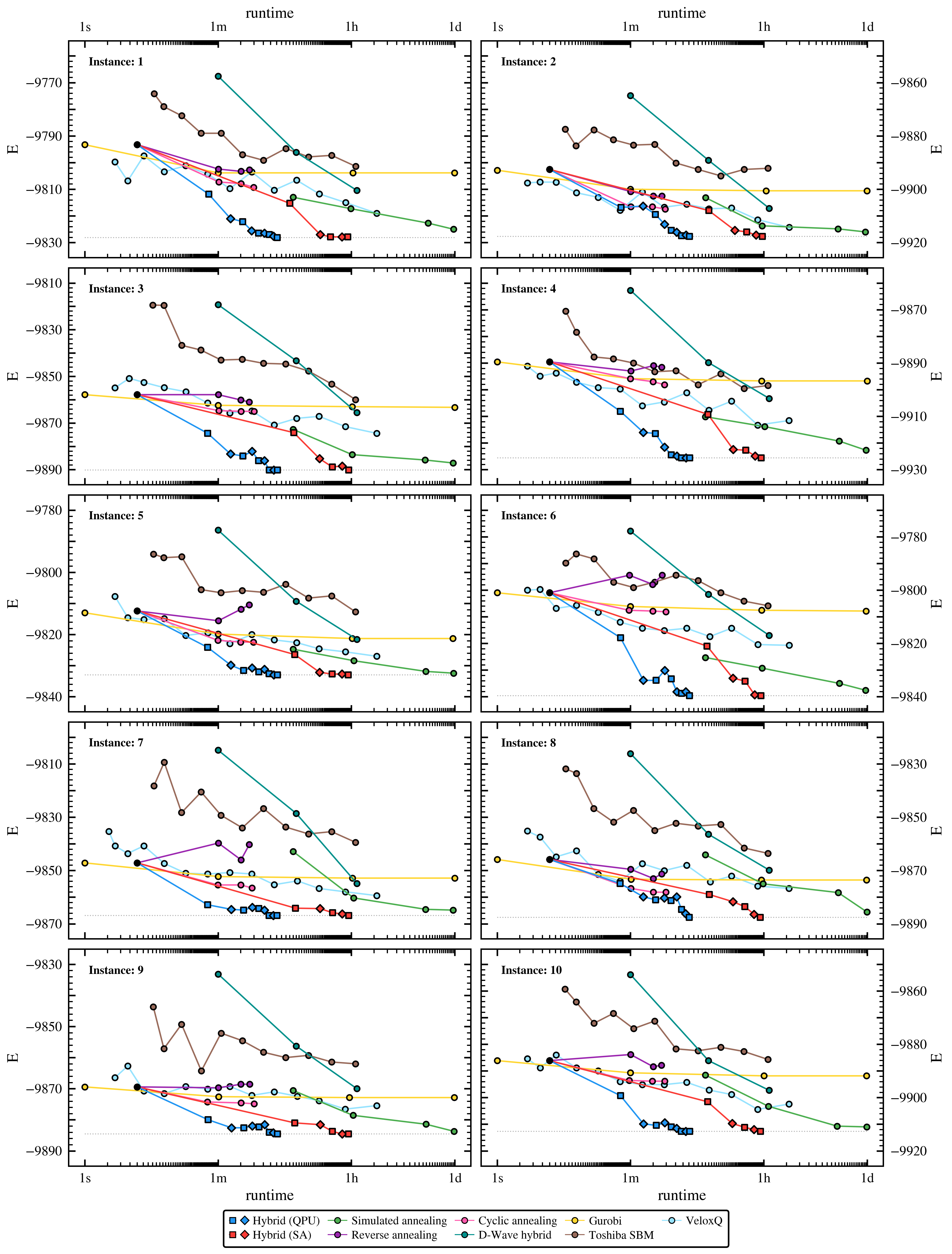}
  \caption{Performance comparison of various quantum and classical solver on $10$ random $5580$-qubit NAT-7 spin-glass instances. The data corresponds to \figref{fig:benchmark_nat7}, but depicts the absolute energies found by each solver.}
  \label{fig:benchmark_nat7_all}
\end{figure*}

Figure~\ref{fig:benchmark_nat7_all} compares the benchmarking results of the investigated quantum and classical solvers in \secref{sec:benchmarks} within the $10$ NAT-7 spin-glass instances. The figure is based on the same data shown in \figref{fig:benchmark_nat7} but highlights the absolute energies found by each solver. Note that the problem instances are designed to feature numerous low-energy local minima, that have energies within a percent of the ground state energy but are separated by large energy barriers ($d_{ham} = (100)$). This allows us to study the solvers' ability to navigate rough energy landscapes. Consequently, the energy differences between solvers are typically small relative to the overall energy scale of the problems, but their corresponding states exhibit a large Hamming distance $d_{ham}$ from one another. The proposed hybrid solver (both QPU and SA sampling), reverse- and cyclic-annealing, and Gurobi use the same initial state $\Psi_i$ (black dot), which is the lowest-energy state obtained from a single forward annealing run on the D-Wave Advantage1 5.4 QPU ($2ms$ annealing time, spin-reversal transformation, $2000$ samples).

\end{document}